\documentclass[amsmath,amssymb,showpacs,showkeywords,twocolumn]{revtex4}
\usepackage{graphicx}
\usepackage{times}
\usepackage{mathrsfs}
\usepackage{amsmath}
\usepackage{graphicx}
\usepackage{epsfig}
\usepackage{dcolumn}
\usepackage{bm}
\usepackage{multirow}
\usepackage{pdfpages}
\usepackage[utf8x]{inputenc}
\DeclareUnicodeCharacter{2212}{-}
\begin{document}
\title{
Nonlinear magnetoelastic wave dynamics and field tunable soliton excitations in hexagonal multiferroic media}

\author{Saumen Acharjee$^{1}$\footnote{saumenacharjee@dibru.ac.in}, 
Kallol Kavas Hazarika$^{2}$\footnote{kallolkavashazarika@gmail.com}, 
Rajneesh Kakoti$^{1}$\footnote{rajneeshkakoti@gmail.com}}
\affiliation{$^{1}$Department of Physics, Dibrugarh University, Dibrugarh 786004, Assam, India}
\affiliation{$^{2}$CSIR-North East Institute of Science and Technology, Jorhat 785006, Assam, India}

\begin{abstract}
We investigate nonlinear magnetoelastic wave dynamics and electrically tunable soliton excitations in hexagonal multiferroic media. By varying the magnetoelastic coupling strength and using a coupled magnetoelastic-ferroelectric continuum model, we found that the system evolves from weakly nonlinear quasiperiodic oscillations to strongly anharmonic yet phase-coherent multimode dynamics. Our results suggest that the dynamics remain bounded and approach distorted limit-cycle behavior rather than chaotic motion despite the enhanced nonlinearity. The excitation spectra and the band dispersion relations reveal that this nonlinear evolution originates from strong magnon-phonon hybridization and coupling-induced renormalization of collective excitation branches, leading to coherent energy exchange among magnetic, elastic, and polarization subsystems. In addition, the coupled dynamics can be reduced to an effective magnetoelastic nonlinear Schr\"{o}dinger equation and support localized excitations such as bright and dark solitons and Kuznetsov-Ma type breathers. Furthermore, it is found that an external electric field modifies both the effective nonlinear coefficient and the dispersion curvature, enabling continuous control of soliton amplitude, width, and stability. The field also induces a saddle-node bifurcation in the magnetization phase space, defining a critical threshold separating multistable and monostable regimes. Our results establish a theoretical framework for electrically tunable nonlinear spin-lattice excitations and soliton engineering in multiferroic systems.
\end{abstract}


\maketitle
\section{Introduction}
Multiferroics are the materials where two or more ferroic orders coexist and interact as a single phase \cite{Schmid1994,Ascher1966,Smolenskii1982}They received significant attention in recent times for their potential application in memory, sensing and information technologies. Consequently, they provide a versatile platform for controlling magnetization, polarization and strain through coupled order parameters~\cite{Spaldin2005,Eerenstein2006,Ramesh2007,Nair2025,Roy2012,Hamou2026,Yu2025}. Since the formal defination of multiferroicity by Hans Schmid~\cite{Schmid1994}, extensive efforts have been given to understand the interplay between spin, lattice, and charge degrees of freedom to realize functionalities beyond conventional ferroic materials~\cite{Fiebig2016,Tokura2014,Disseler2015,Tian2016}. A central challenge is to realize strong magnetoelectric coupling while maintaining ferroic orders at experimentally relevant temperatures and fields \cite{Khomskii2009}. Type-I multiferroics, such as BiFeO$_3$, exhibit large polarization and high ordering temperatures but have typically weak coupling between magnetic and ferroelectric orders \cite{Catalan2009,CarranzaCelis2019,Long2024}, whereas type-II multiferroics display intrinsically strong coupling at the expense of smaller polarizations and lower transition temperatures \cite{Cheong2007,Zhang2025,Mostovoy2024}. Thus the basic objective is to combine the stability of type-I systems with the strong coupling characteristic of type-II multiferroics. This motivates the search for new structural classes of multiferroic materials that combine both characteristics within a single system~\cite{Khomskii2009}.

Hexagonal rare earth manganites $\mathcal{R}$MnO$_3$ ($\mathcal{R}$ = Y, Ho, Er, etc.) provide a distinctive platform for exploring coupled ferroic excitations \cite{Ye2015,Petit2007,Park2020,Kalashnikova2003,Curnoe2006,Huang2015}. In these systems, improper ferroelectricity emerges from a geometric trimerization of MnO$_5$ bipyramids, resulting in strong polarization with Curie temperatures exceeding $10^3$~K~\cite{Nordlander2019,Aken2004,Nordlander2022}. The coexistence of ferroelectric and antiferromagnetic orders, combined with topologically protected vortex-antivortex domain structures and functional domain walls, provides a rich landscape of coupled spin, lattice and polarization phenomena~\cite{Giraldo2021,Lin2020,Wu2024}. Importantly, the separation of ferroelectric and magnetic ordering temperatures allows relatively independent manipulation of polarization and spin dynamics. Thus this offer hybrid excitations~\cite{Mostovoy2024} and field tunable responses in multiferroic media~\cite{Bai2021,Sagayama2008,Pradhan2020}.

A particularly relevant class of hybrid collective excitations are the Magnetoelastic waves (MEWs) which arise from the coherent hybridization of spin waves and elastic phonons~\cite{Park2020}. They facilitate the transfer of energy and information between magnetic and mechanical degrees of freedom~\cite{Lejman2019,Dreher2012,Streib2019,Hashimoto2018}. So, MEWs have been widely explored in ferromagnets and piezoelectric–ferromagnetic heterostructures, where strain-mediated coupling allows controllable magnon propagation and enable reconfigurable magnonic functionalities~\cite{Azovtsev2023,Yamamoto2023,Hashimoto2018,Burdin2021,Esteras2022,Li2025}. In multiferroics, however, an additional dynamical degree of freedom, polarization can directly modulate the wave propagation along with spin and strain. This opens the possibility of threefold hybridization among magnons, phonons, and polariton modes, allowing external electric and magnetic fields to tune not only dispersion relations but also the coupling strengths among collective excitations~\cite{Mostovoy2006,Pimenov2006,Katsura2007}. Although spin and lattice interactions have been widely explored but a systematic theoretical description of MEWs in multiferroic media that treats polarization dynamics on the same footing as spin and lattice degrees of freedom remains largely unexplored. 

Beyond the linear regime, nonlinear interactions between coupled excitations can lead to the formation of solitons, which are stable localized wavepackets sustained by the balance between nonlinearity and dispersion \cite{Scott2003,Copie2025,Kartashov2011,Iqbal2025}. Solitons play a central role in all fields of nonlinear physics and condensed matter physics, where they function as robust carriers of information that are resilient to scattering and dispersion \cite{Shaukat2022}. In magnetic systems, nonlinear excitations in terms of domain walls, magnetic solitons, and breathers have been extensively explored in ferromagnets and engineered heterostructures~\cite{Kim2014,Kosevich1990,Sievers1988}. Multiferroics offer a qualitatively new range of possibilities for nonlinear excitations in terms of hybrid solitons whose characteristics can be controlled using external electric and magnetic fields \cite{Katsura2007}. Accordingly, we examine MEWs in hexagonal multiferroics involving magnetic, elastic, and ferroelectric degrees of freedom. In the linear regime, we find that MEWs exhibit a three-way hybridization between magnonic, phononic, and ferroelectric excitations and show that these hybridization gaps can be controlled using external electric and magnetic fields \cite{Charalampidis2015}. In the nonlinear regime, a multiple-scale method provides a nonlinear Schr\"{o}dinger equation supporting bright, dark, and breather-type solitons~\cite{Chowdury2017,Meng2012,Charalampidis2015}. 

The novelty of this work lies in three aspects: (i) a unified treatment of MEWs in hexagonal multiferroics that explicitly incorporates polarization dynamics on the same footing as spin and lattice degrees of freedom, (ii) demonstration of field-tunable hybridization and dispersion in the linear collective mode spectrum, and (iii) prediction and classification of nonlinear soliton excitations in hexagonal multiferroic media. 

The paper is organized as follows. In Sec.~II we introduce the model Hamiltonian and obtain the coupled equation of motion for MEWs in hexagonal multiferroic system.  The magnetoelastic dispersion are discussed in In Sec.~III. In Sec.~IV, we reduce the coupled dynamics to a nonlinear Schr\"{o}dinger equation and analyze the corresponding envelope solitons. We present the soliton solutions in Sec.~V. Finally, Sec.~VI summarizes our results.

\section{Coupled Equation of motion for hexagonal multiferroic system}
We consider a noncentrosymmetric hexagonal multiferroic belongs to the $6mm$ point group relevant for hexagonal manganites $\mathcal{R}$MnO$_3$ ($\mathcal{R}=$ Y, Ho, Er, etc.)~\cite{Ye2015,Petit2007,Park2020,Kalashnikova2003,Curnoe2006,Huang2015}. The lack of inversion symmetry in this class allow linear magnetoelectric and strain–polarization couplings in the free energy. The total Hamiltonian density of the system can be written as~\cite{Born1954,Nye1985,Landau1986}
\begin{equation}
\mathcal{H} = \mathcal{H}_\text{elas} + \mathcal{H}_\text{mag}+ \mathcal{H}_\text{mag-elas}  + \mathcal{H}_\text{mag-elec},
\label{eq1}
\end{equation}
where $\mathcal{H}_\text{elas}$, $\mathcal{H}_\text{mag}$, $\mathcal{H}_\text{mag-elas}$, and $\mathcal{H}_\text{mag-elec}$ denote the elastic, magnetic, magnetoelastic, and magnetoelectric energy densities, respectively. The energy contributions are expressed in terms of the lowest-order invariants compatible with $6mm$ symmetry. This symmetry condition determines the independent elastic constants and limits the allowed couplings between strain, magnetization, and polarization. In contrast, centrosymmetric hexagonal crystals of the $6/mmm$ class would forbid linear magnetoelectric terms of the form $\mathbf{m}\cdot\mathbf{p}$ ~\cite{Fiebig2016}. The elastic energy density $\mathcal{H}_{\text{elas}}$ of a hexagonal crystal is given by ~\cite{Born1954,Nye1985,Landau1986,Mouhat2014}.
\begin{equation}
\begin{aligned}
\mathcal{H}_{\text{elas}} = & \ \tfrac{1}{2} \mathcal{C}_{11} \left(\epsilon_{xx}^2 + \epsilon_{yy}^2 \right) + \mathcal{C}_{12}\,\epsilon_{xx}\epsilon_{yy} + \tfrac{1}{2} \mathcal{C}_{33}\,\epsilon_{zz}^2 \\
& + \mathcal{C}_{13} (\epsilon_{xx} + \epsilon_{yy})\epsilon_{zz} + 2\mathcal{C}_{44}\,(\epsilon_{yz}^2 + \epsilon_{zx}^2) 
\\& + \mathcal{C}_{66}\,\epsilon_{xy}^2 ,
\end{aligned}
\label{eq2}
\end{equation}
where $\mathcal{C}_{11}, \mathcal{C}_{12}, \mathcal{C}_{33}, \mathcal{C}_{13}, \mathcal{C}_{44}, \mathcal{C}_{66}$ are the elastic stiffness constants arising from the hexagonal symmetry of the multiferroic media, and $\epsilon_{ij}$ are the strain components. For hexagonal symmetry, the elastic constants satisfy $C_{66}=(C_{11}-C_{12})/2$. But we retain $C_{66}$ explicitly for notational convenience. In linear elasticity, the strain tensor is related to the displacement field $\mathbf{u} = (u_x,u_y,u_z)$ as
\begin{equation}
\epsilon_{ij} = \tfrac{1}{2} \left(\frac{\partial u_i}{\partial x_j} + \frac{\partial u_j}{\partial x_i}\right).
\label{eq3}
\end{equation}

The Lagrangian density for the elastic system can be defined as
\begin{equation}
\mathcal{L} = \tfrac{1}{2} \rho \left(\dot{u}_x^2 + \dot{u}_y^2 + \dot{u}_z^2 \right) - \mathcal{H}_{\text{elas}},
\label{eq4}
\end{equation}
where $\rho$ is the mass density.

The Euler–Lagrange equations for each displacement component $u_i$ can be written as
\begin{equation}
\frac{\partial}{\partial t}\left(\frac{\partial \mathcal{L}}{\partial \dot{u}_i}\right) 
- \sum_j \frac{\partial}{\partial x_j}\left(\frac{\partial \mathcal{L}}{\partial (\partial_{x_j} u_i)}\right) = 0.
\label{eq5}
\end{equation}

The magnetoelastic and magnetoelectric couplings are constructed from the lowest order scalar invariants of $\epsilon_{ij}$, $\mathbf{m}$, and $\mathbf{p}$ allowed by $6mm$ symmetry. Only bilinear terms that transform as the identity representation are retained, while higher order invariants are neglected in the weak-coupling, long-wavelength limit. Accordingly the magnetoelastic free energy density is~\cite{Fiebig2016}
\begin{align}
\notag
\mathcal{F}_\text{mag-elas} = B_1(m_x^2\epsilon_{xx} + m_y^2\epsilon_{yy} + m_x m_y \epsilon_{xy}) 
- B_2 m_z^2 \epsilon_{zz} 
\\- B_3 m_z^2 (\epsilon_{xx}+\epsilon_{yy}) 
+ B_4(m_y m_z \epsilon_{yz} + m_x m_z \epsilon_{zx}),
\label{eq6}
\end{align}
where $B_1,B_2,B_3,B_4$ are magnetoelastic coupling constants arising from hexagonal symmetry and quantify the coupling between magnetization and strain.

Thus the total stress tensor can be obtained by considering both elastic and magnetoelastic contributions and can be defined as
\begin{equation}
\sigma_{ij}^{\text{total}} = \frac{\partial \mathcal H_{\rm elas}}{\partial\epsilon_{ij}} + \frac{\partial \mathcal{F}_\text{mag-elas}}{\partial\epsilon_{ij}}
\label{eq7}
\end{equation}
The stress tensor components using hexagonal symmetry can be written as
\[
\begin{aligned}
\sigma_{xx} &= \mathcal{C}_{11}\epsilon_{xx} + \mathcal{C}_{12}\epsilon_{yy} + \mathcal{C}_{13}\epsilon_{zz}
+ B_1 m_x^2 - B_3 m_z^2,\\
\sigma_{yy} &= \mathcal{C}_{11}\epsilon_{yy} + \mathcal{C}_{12}\epsilon_{xx} + \mathcal{C}_{13}\epsilon_{zz}
+ B_1 m_y^2 - B_3 m_z^2,\\
\sigma_{zz} &= \mathcal{C}_{33}\epsilon_{zz} + \mathcal{C}_{13}(\epsilon_{xx}+\epsilon_{yy}) - B_2 m_z^2,\\
\sigma_{xy} &= 2\mathcal{C}_{66}\epsilon_{xy} + \tfrac{1}{2}B_1 m_x m_y,\\
\sigma_{yz} &= 2\mathcal{C}_{44}\epsilon_{yz} + \tfrac{1}{2}B_4 m_y m_z,\\
\sigma_{zx} &= 2\mathcal{C}_{44}\epsilon_{zx} + \tfrac{1}{2}B_4 m_x m_z.
\end{aligned}
\]
In Voigt notation, the stress–strain relation can be written as
\begin{equation}
\begin{pmatrix}
\sigma_{xx} \\ \sigma_{yy} \\ \sigma_{zz} \\ \sigma_{yz} \\ \sigma_{zx} \\ \sigma_{xy}
\end{pmatrix}
=
\begin{pmatrix}
\mathcal{Q}_1 & \mathcal{C}_{12} & \mathcal{C}_{13} & 0 & 0 & 0 \\
\mathcal{C}_{12} & \mathcal{Q}_2 & \mathcal{C}_{13} & 0 & 0 & 0 \\
\mathcal{C}_{13} & \mathcal{C}_{13} & \mathcal{Q}_3 & 0 & 0 & 0 \\
0 & 0 & 0 & \mathcal{Q}_4 & 0 & 0 \\
0 & 0 & 0 & 0 & \mathcal{Q}_5 & 0 \\
0 & 0 & 0 & 0 & 0 & \mathcal{Q}_6
\end{pmatrix}
\begin{pmatrix}
\epsilon_{xx} \\ \epsilon_{yy} \\ \epsilon_{zz} \\ \epsilon_{yz} \\ \epsilon_{zx} \\ \epsilon_{xy}
\end{pmatrix}.
\label{eq8}
\end{equation}
where, $\mathcal{Q}_1 = \mathcal{C}_{11}+B_1m_x^2 -B_3m_z^2$, $\mathcal{Q}_2 = \mathcal{C}_{11}+B_1m_y^2-B_3m_z^2$, $\mathcal{Q}_3 = \mathcal{C}_{33}-B_2m_z^2$, $\mathcal{Q}_4 = 2\mathcal{C}_{44}+B_4m_ym_z$, $\mathcal{Q}_5 = 2\mathcal{C}_{44}+B_4m_xm_z$ and $\mathcal{Q}_6 = 2\mathcal{C}_{66}+B_1m_xm_y$\\

The Hamiltonian for the magnetic subsystem in the continuum limit can be given by~\cite{Landau1986}
\begin{equation}
\mathcal{H}_\text{mag} = \frac{1}{2}\mathcal{A} \, (\nabla \mathbf{m})^2 
+ \frac{1}{2}\mathcal{K}_u \, (\mathbf{m} \cdot \hat{\mathbf{n}})^2 - \mu_0 \mathbf{H} \cdot \mathbf{m},
\label{eq9}
\end{equation}
where, the first term corresponds to the exchange interaction with $\mathcal{A}$ being the exchange stiffness constant. The second term correspond to the uniaxial anisotropy with $\mathcal{K}_u$ is the uniaxial anisotropy constant along the easy axis $\hat{\mathbf{n}}$. The term $\mu_0 \mathbf{H}$ correspond to the Zeeman energy. We introduce the normalized magnetization $\mathbf{m}=\mathbf{M}/M_s$ with $|\mathbf{m}|=1$, where $M_s$ is the saturation magnetization.

The effective field $\mathbf{H}_{\text{eff}}$ can be obtained using the variational derivative of Eq.(\ref{eq9})
\begin{equation}
\mathbf{H}_{\text{eff}} = -\frac{\delta \mathcal{H}_\text{mag}}{\delta \mathbf{m}} = -\mathcal{A} \nabla^2 \mathbf{m} - \mathcal{K}_u m_z \hat{z}+ \mu_0 \mathbf{H}
\label{eq10}
\end{equation}

So, the magnetization dynamics is governed by the Landau–Lifshitz–Gilbert (LLG) equation
\begin{equation}
\dot{\mathbf{m}} = -\gamma \mathbf{m}\times \mathbf{H}_{\text{eff}} + \alpha \mathbf{m}\times \dot{\mathbf{m}}+ \mathbf{T}_\text{mag-elas},
\label{eq11}
\end{equation}
where, $\gamma$ is the gyromagnetic ratio and $\alpha$ is the Gilbert damping constant. The term \(\mathbf{T}_\text{mag-elas}\) characterize the magnetoelastic torque of the system
\[
\mathbf{T}_\text{mag-elas}
= -\gamma\,\mathbf{m}\times
\left(
\frac{\delta\mathcal{H}_\text{mag-elas}}{\delta\mathbf{m}}
\right).
\]
The electric energy density discussed in Eq.~(\ref{eq1}) is defined as
\begin{equation}
\mathcal{H}_{\text{elec}} = \frac{\kappa}{2}(p_x^2+p_y^2) + \tfrac{\delta}{2}(\nabla\mathbf p)^2- \mathbf{E}\cdot \mathbf{p},
\label{eq12}
\end{equation}
where $\kappa = 1/\chi_e$ represents the intrinsic dielectric stiffness. The gradient term penalizes polarization inhomogeneities essential for describing polarization textures and propagating hybrid modes.

The magnetoelectric energy density discussed in Eq.(\ref{eq1}) for the hexagonal multiferroic system can be defined as~\cite{Landau1986}
\begin{equation}
\mathcal{H}_\text{mag-elec} = \gamma_1 (\epsilon_{xx} p_x + \epsilon_{yy} p_y) + \gamma_2(m_x p_x + m_y p_y).
\label{eq13}
\end{equation}
where, \(\gamma_1\) and \(\gamma_2\) represents the electrostrictive and magnetoelectric coupling strengths. 
The first term  links lattice strain to local polarization, 
providing a direct route for strain-controlled electric responses. The second term represents the 
intrinsic magnetoelectric interaction through which magnetic order influences polarization. For $6mm$ symmetry, the leading magnetoelectric invariant is linear in both $\mathbf{m}$ and $\mathbf{p}$, yielding Eq.~(\ref{eq13}); tensorial anisotropies allowed by $6mm$ symmetry are neglected for simplicity. Such linear magnetoelectric terms are forbidden in the centrosymmetric $6/mmm$ class.

The polarization dynamics is governed by the Landau-Khalatnikov equation~\cite{Maslovskaya2021}
\begin{equation}
k \ddot{p}_i + \Gamma \dot{p}_i = -\frac{\delta \mathcal{H}}{\delta p_i},
\label{eq14}
\end{equation}
where the inertial term describes high-frequency optical phonon dynamics; the overdamped limit $k\rightarrow0$ recovers the standard relaxational form.
Thus, in view of Eqs. (\ref{eq5}) - (\ref{eq14}), the coupled magnetoelastic equations can be written as

\begin{align}
\label{eq15}
\rho \ddot{u}_x &= 
(\mathcal{C}_{11}\partial_x^2 
+ \mathcal{C}_{66}\partial_y^2 
+ 2\mathcal{C}_{44}\partial_z^2)\,u_x 
+ (\mathcal{C}_{12}+\mathcal{C}_{66})\,\partial_x\partial_y u_y
\notag\\
&\quad + (\mathcal{C}_{13}+2\mathcal{C}_{44})\,\partial_x\partial_z u_z 
+ \partial_x(B_1 m_x^2 - B_3 m_z^2 - \gamma_1 p_x)
\notag\\
&\quad + \partial_y(B_1 m_x m_y)
+ \partial_z(B_4 m_x m_z),
\\[1.0em]
\label{eq16}
\rho \ddot{u}_y &=
(\mathcal{C}_{66}\partial_x^2 
+ \mathcal{C}_{11}\partial_y^2 
+ 2\mathcal{C}_{44}\partial_z^2)\,u_y
+ (\mathcal{C}_{12}+\mathcal{C}_{66})\,\partial_x\partial_y u_x
\notag\\
&\quad + (\mathcal{C}_{13}+2\mathcal{C}_{44})\,\partial_y\partial_z u_z
+ \partial_y(B_1 m_y^2 - B_3 m_z^2 - \gamma_1 p_y)
\notag\\
&\quad + \partial_x(B_1 m_x m_y)
+ \partial_z(B_4 m_y m_z),
\\[1.0em]
\label{eq17}
\rho \ddot{u}_z &=
(\mathcal{C}_{44}\partial_x^2 
+ \mathcal{C}_{44}\partial_y^2 
+ \mathcal{C}_{33}\partial_z^2)\,u_z
+ \mathcal{C}_{13}\,\partial_z(\partial_x u_x + \partial_y u_y)
\notag\\
&\quad
+ \partial_z(-B_2 m_z^2)
+ \partial_x(B_4 m_x m_z)
+ \partial_y(B_4 m_y m_z),
\\
\label{eq18}
\dot{m}_x &=  \alpha(m_y \dot{m}_z 
- m_z \dot{m}_y)
-\gamma\Big[
m_y\Big\{\mathcal{A}\nabla^2 m_z + K_u m_z 
\notag\\
&\quad- \mu_0 H_z
+ 2B_2 m_z \partial_z u_z
 + 2B_3 m_z(\partial_x u_x + \partial_y u_y)
\notag\\
&\quad- \tfrac{1}{2}B_4\Big(m_x (\partial_x u_z+\partial_z u_x)
 + m_y (\partial_z u_y+\partial_y u_z)\Big)
\notag\\
&\quad - \gamma_2 p_z
\Big\}
- m_z\Big\{\mathcal{A}\nabla^2 m_y 
 - \mu_0 H_y
- 2B_1 m_y \partial_y u_y
\notag\\
&\quad - \tfrac{1}{2}B_1 m_x (\partial_y u_x +\partial_x u_y)
- \tfrac{1}{2}B_4 m_z (\partial_z u_y+\partial_y u_z)
\notag\\
&\quad- \gamma_2 p_y
\Big\}
\Big],
\\
\label{eq19}
\dot{m}_y &= \alpha(m_z \dot{m}_x - m_x \dot{m}_z)-\gamma\Big[
m_z\Big\{\mathcal{A}\nabla^2 m_x - \mu_0 H_x
\notag\\
&\quad- 2B_1 m_x \partial_x u_x
- \tfrac{1}{2}B_1 m_y (\partial_y u_x
+\partial_x u_y)
- \gamma_2 p_x 
\notag\\
&\quad- \tfrac{1}{2} B_4 m_z (\partial_x u_z
+\partial_z u_x)
\Big\}- m_x\Big\{\mathcal{A}\nabla^2 m_z 
+ K_u m_z \notag\\
&\quad - \mu_0 H_z
+ 2B_2 m_z \partial_z u_z
+ 2B_3 m_z(\partial_x u_x+\partial_y u_y)
\notag\\
&\quad
- \tfrac{1}{2}B_4\Big(m_x (\partial_x u_z+\partial_z u_x)
+ m_y (\partial_z u_y+\partial_y u_z)\Big)
\notag\\
&\quad
- \gamma_2 p_z
\Big\}\Big]
,
\\
\label{eq20}
\dot p_x &= -\Gamma\Big\{(\kappa - \delta\,\nabla^2) p_x + \gamma_1\,\partial_x u_x + \gamma_2\,m_x- E_x 
\Big\}\\
\label{eq21}
\dot p_y &= -\Gamma\Big\{(\kappa- \delta\,\nabla^2) p_y  + \gamma_1\,\partial_y u_y + \gamma_2\,m_y
- E_y\Big\}
\end{align}
These coupled Eqs.(\ref{eq15}) - (\ref{eq21}) describe nonlinear MEWs whose balance between dispersion and nonlinearity enables field-tunable soliton excitations in hexagonal multiferroics.
\begin{figure*}[t]
\centerline
\centerline{ 
\hspace{-0.5mm}
\includegraphics[scale=0.15]{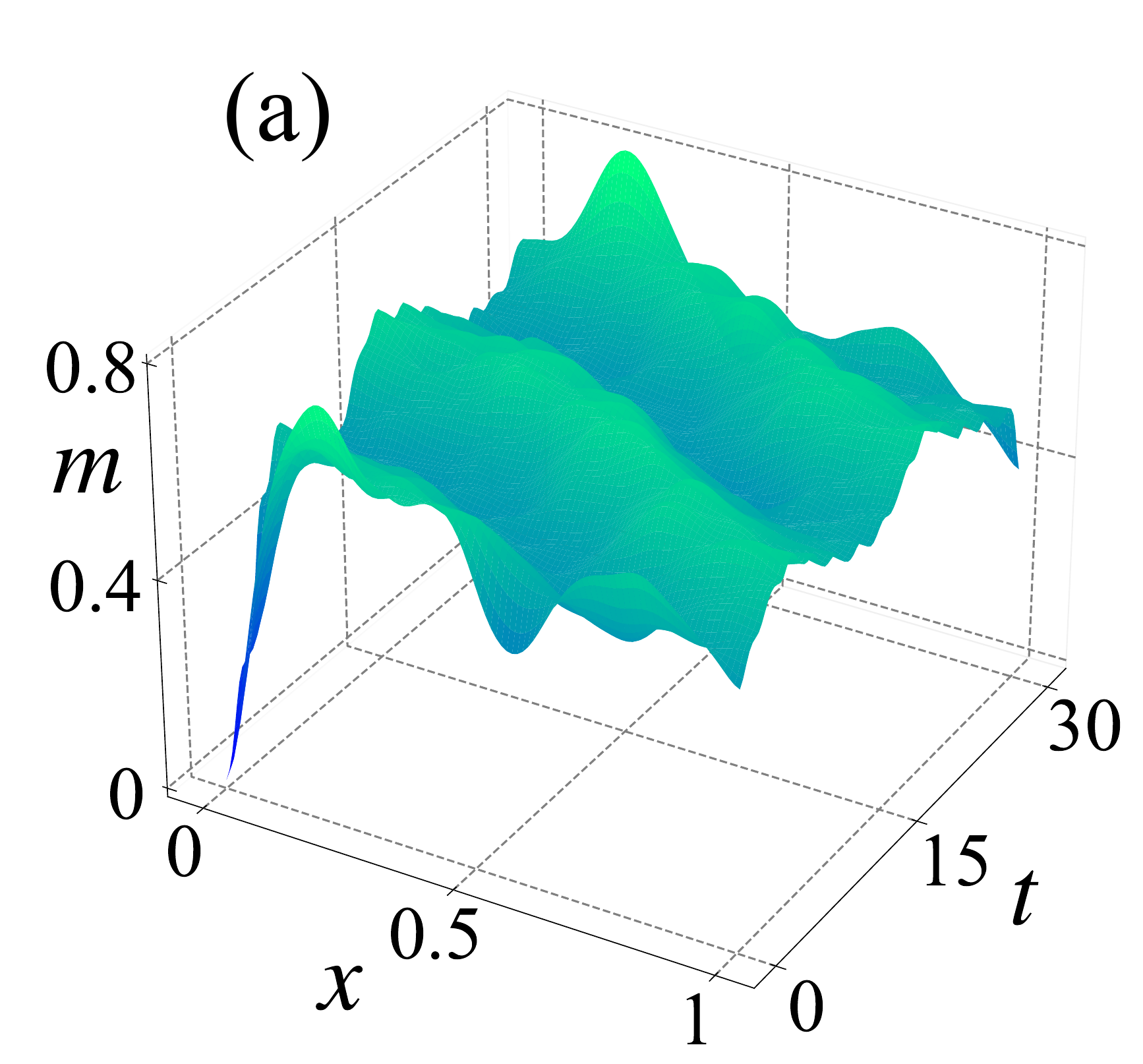}
\includegraphics[scale=0.23]{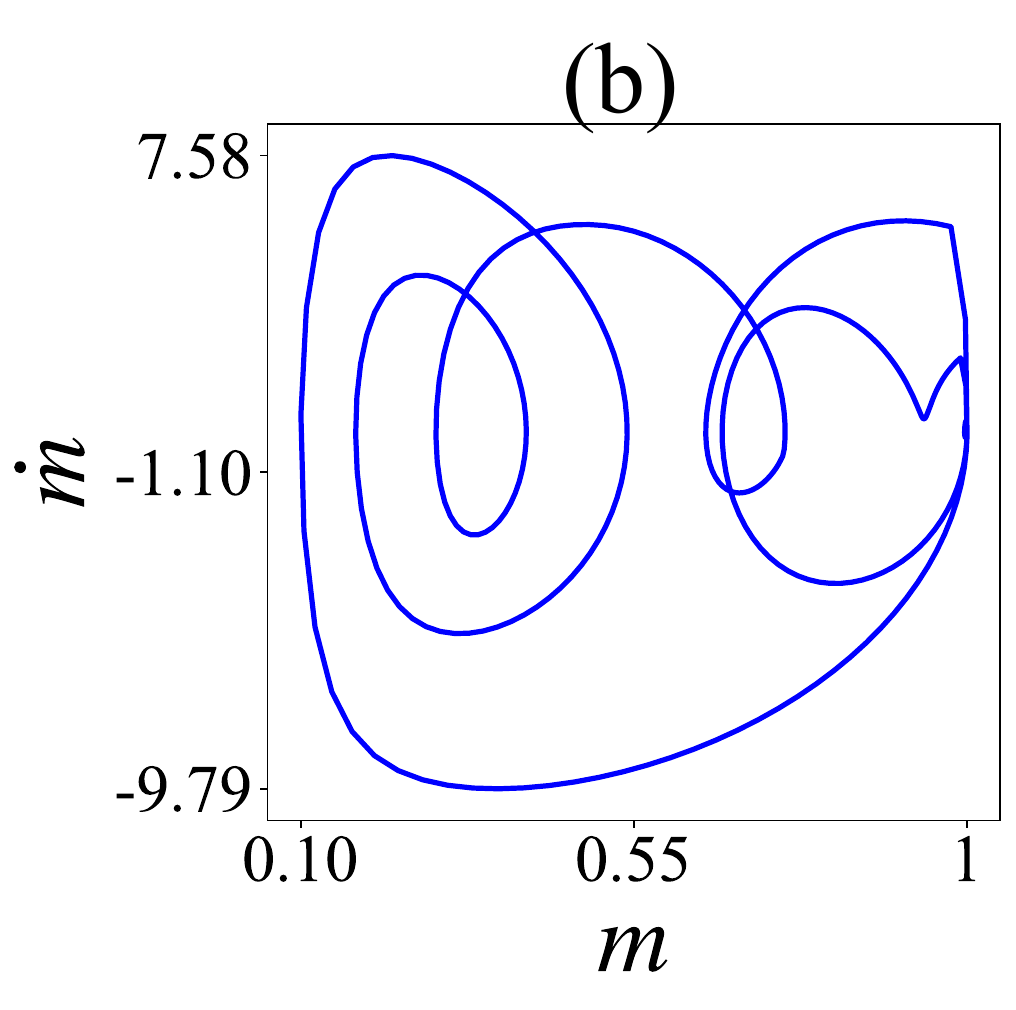}
\hspace{1.5mm}
\includegraphics[scale=0.155]{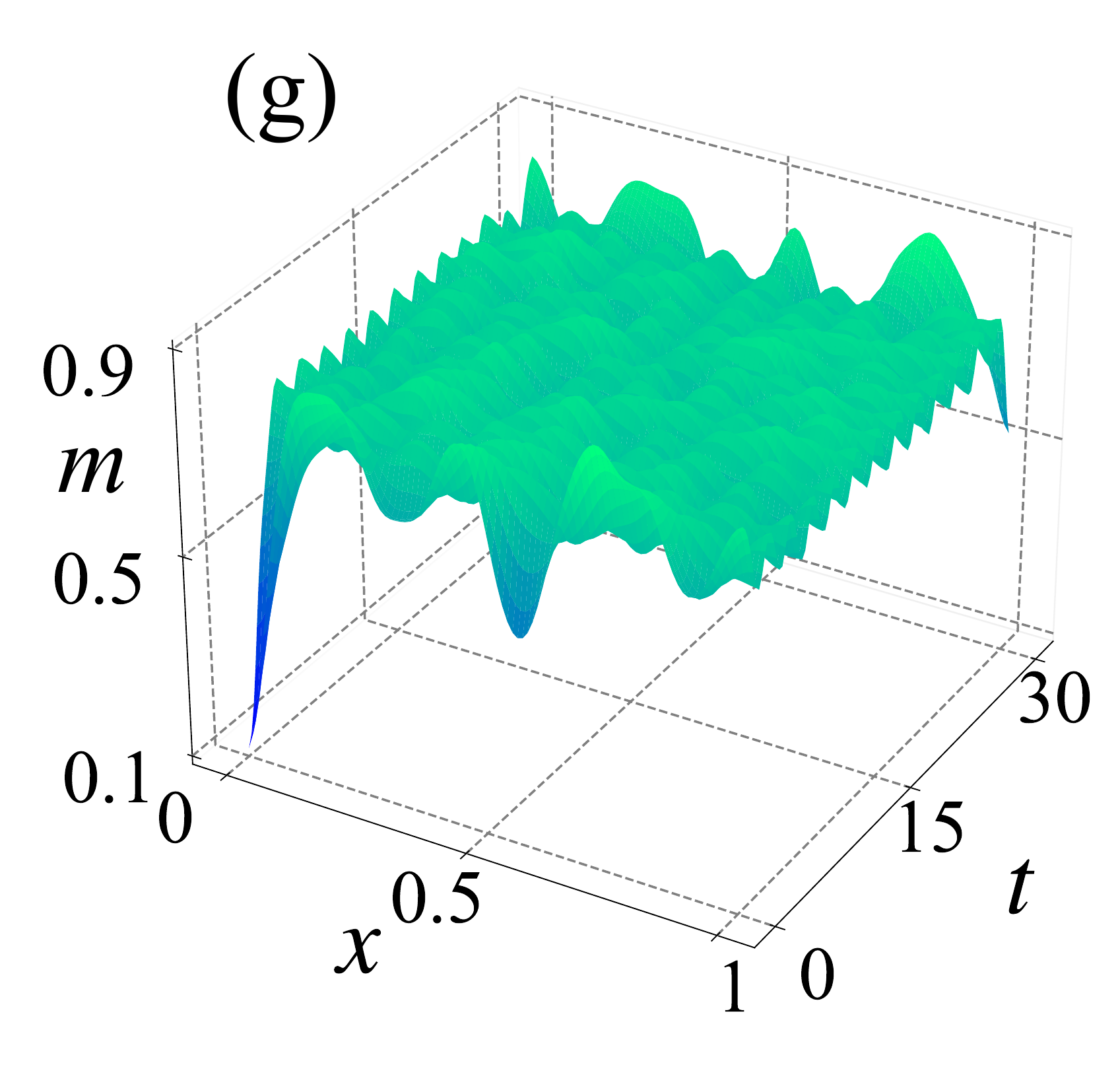}
\includegraphics[scale=0.23]{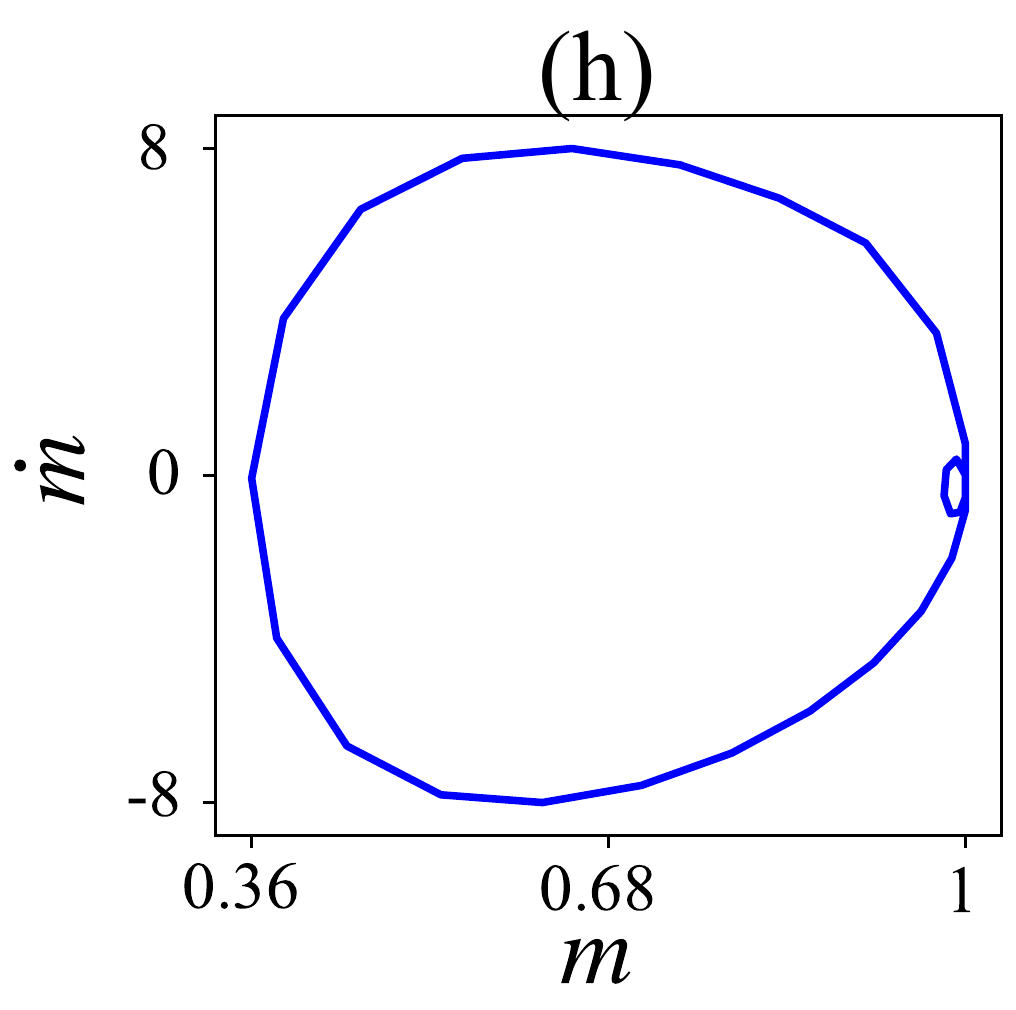}}
\centerline
\centerline{ 
\includegraphics[scale=0.15]{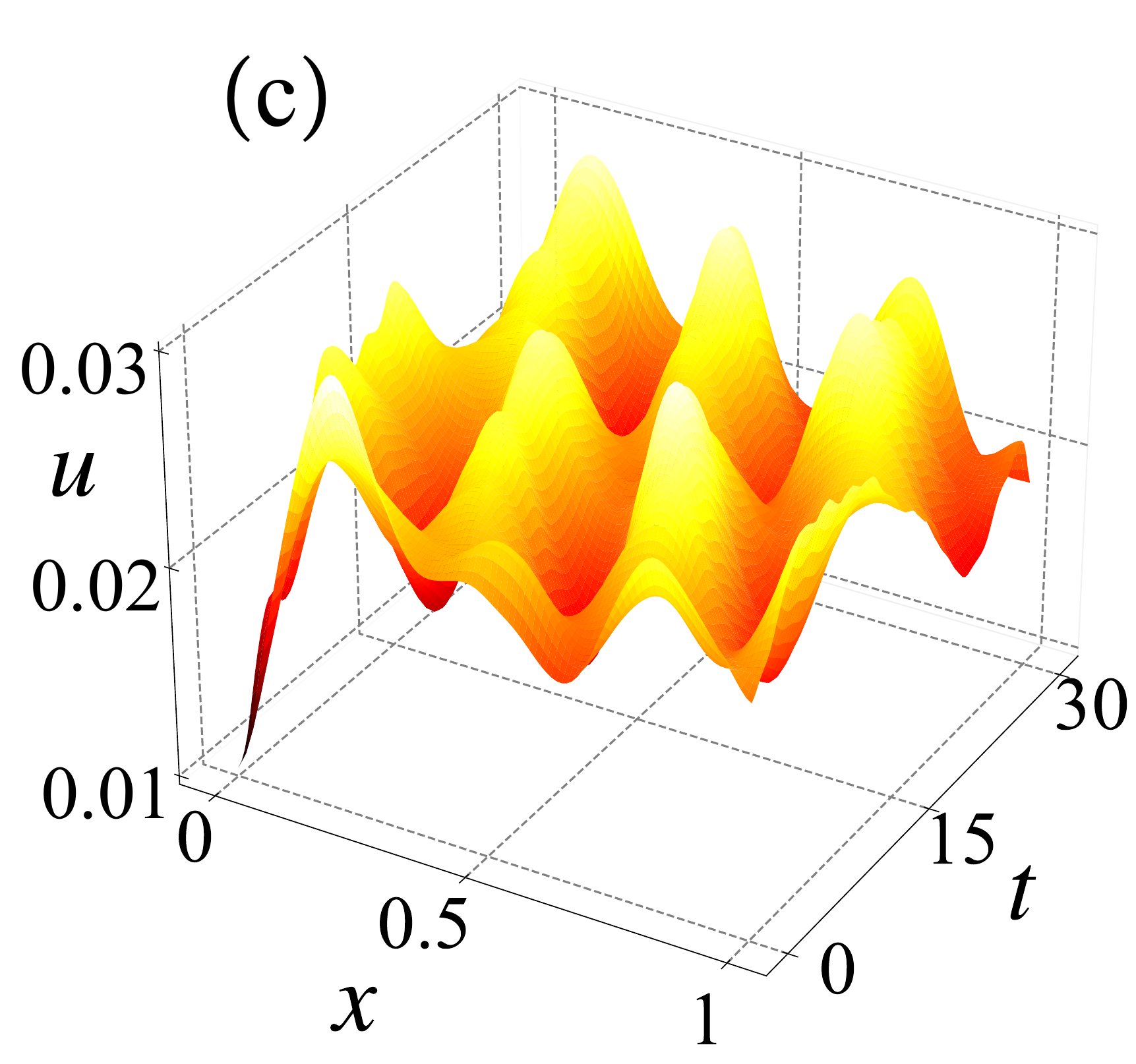}
\includegraphics[scale=0.23]{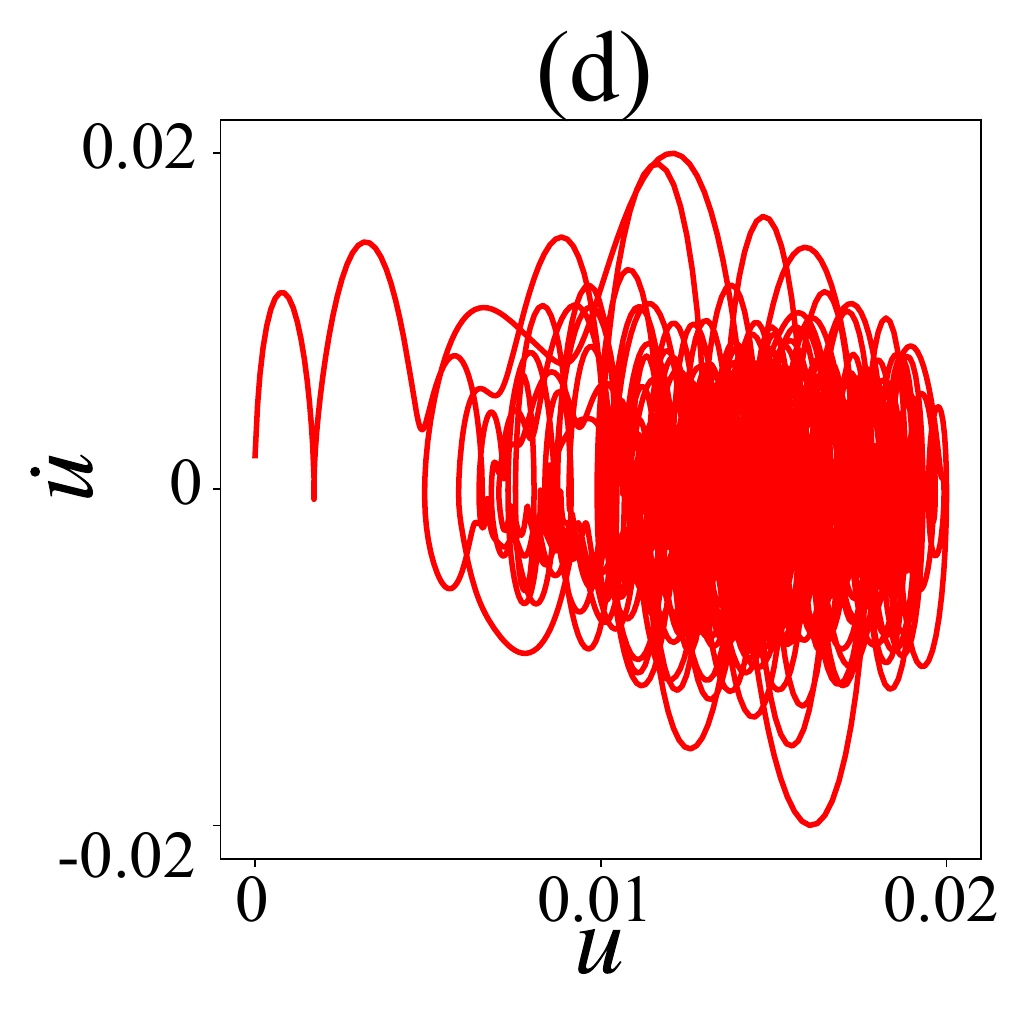}
\includegraphics[scale=0.155]{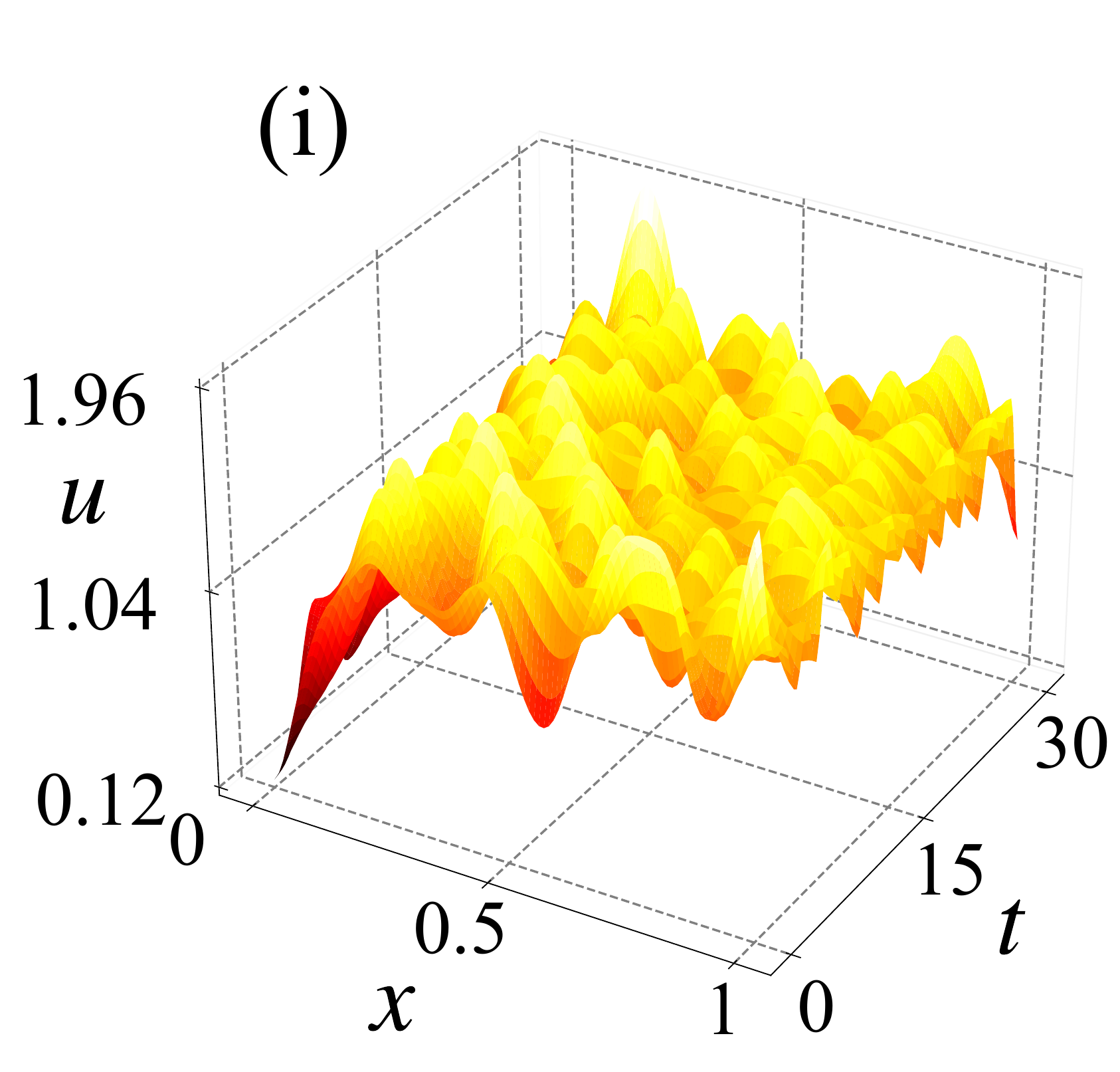}
\includegraphics[scale=0.23]{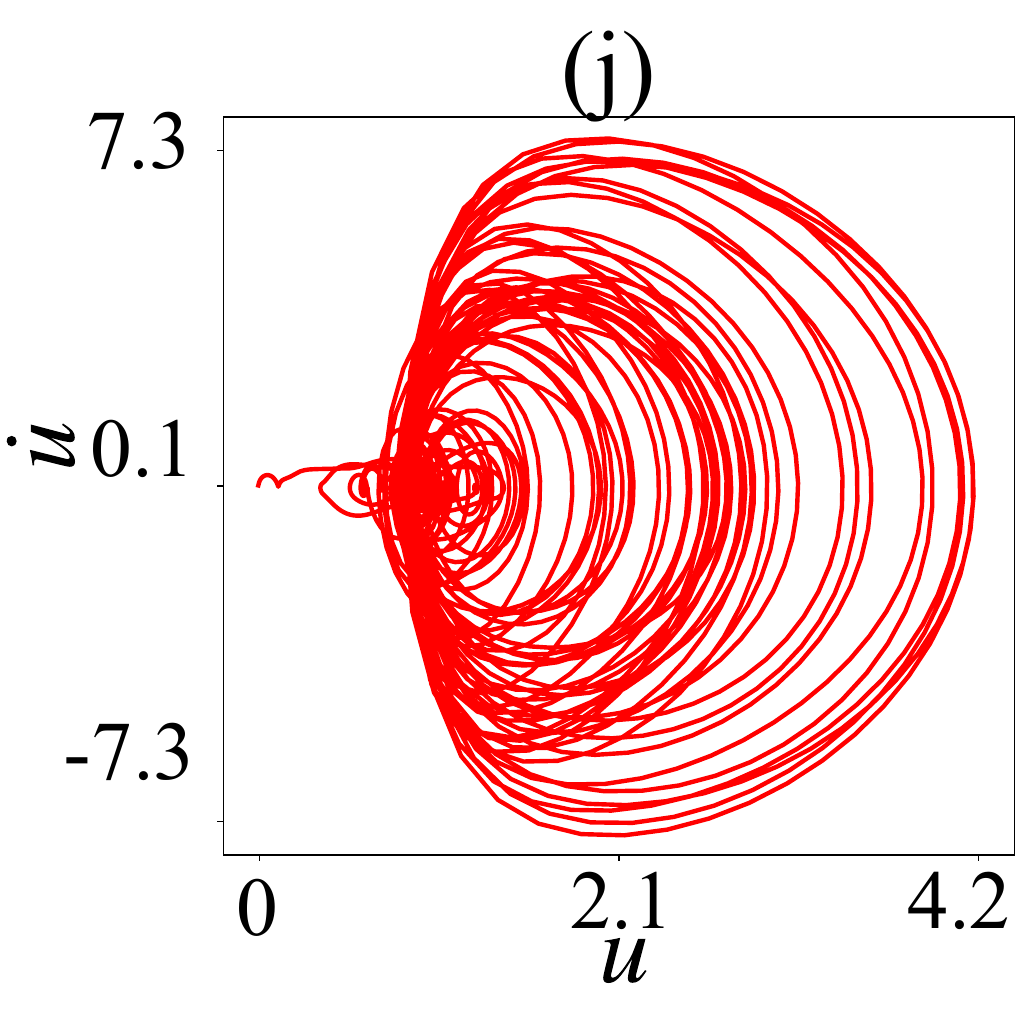}}
\centerline
\centerline{ 
\hspace{-0.5mm}
\includegraphics[scale=0.15]{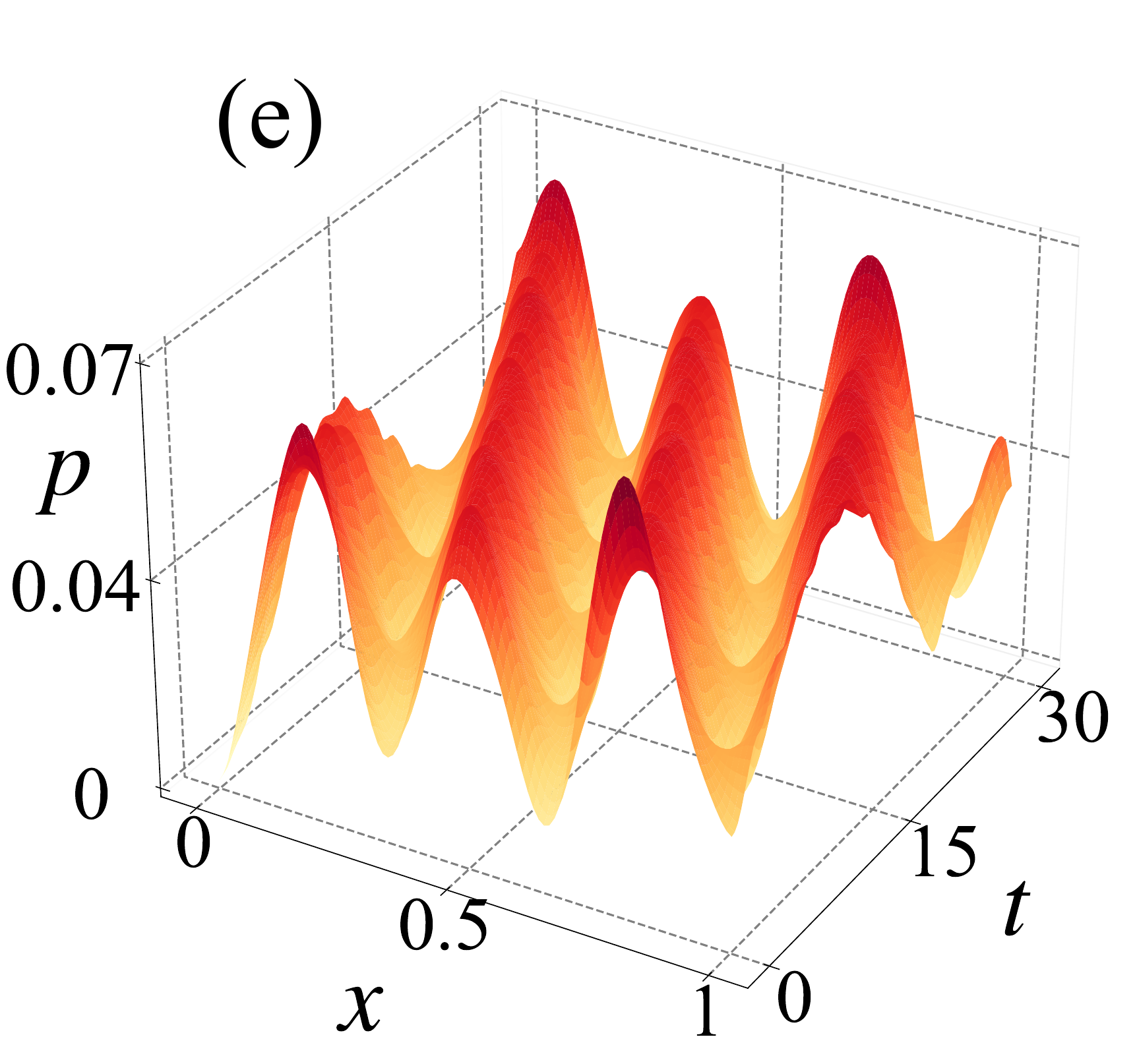}
\includegraphics[scale=0.24]{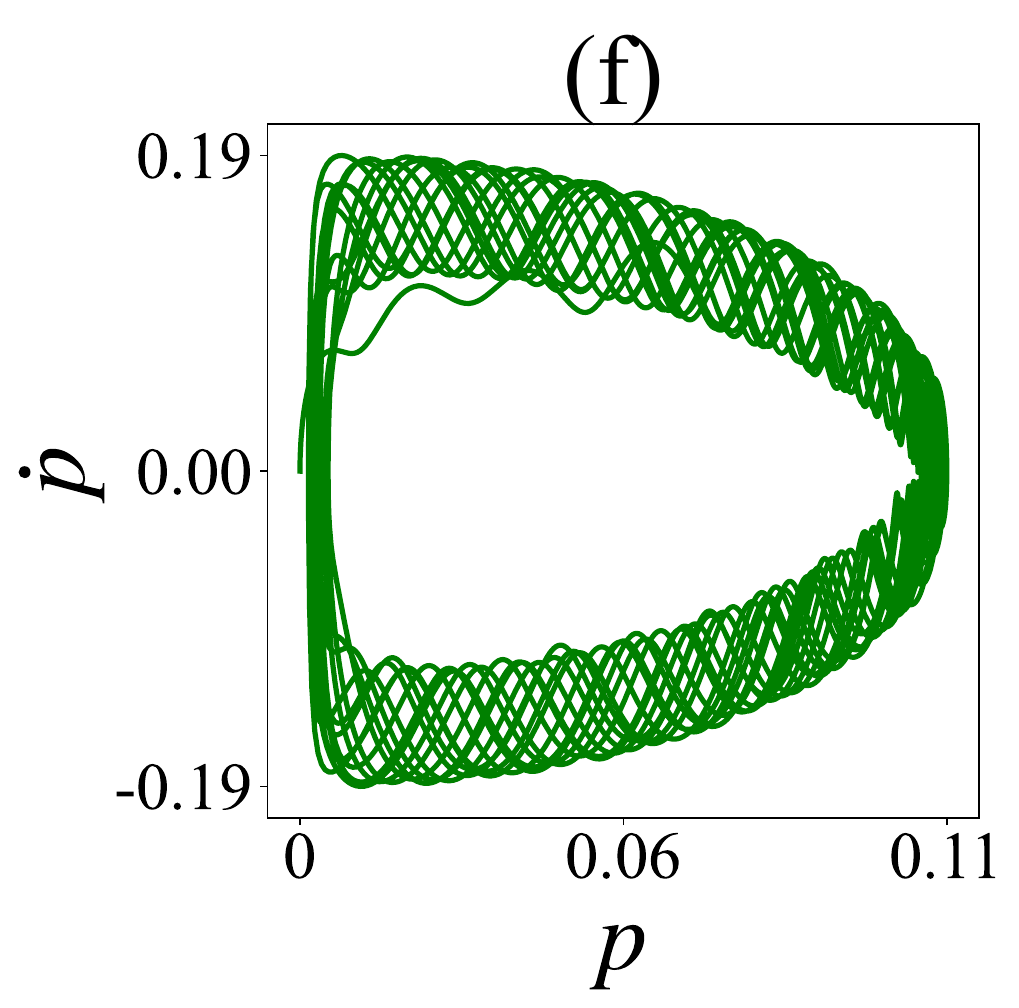}
\includegraphics[scale=0.16]{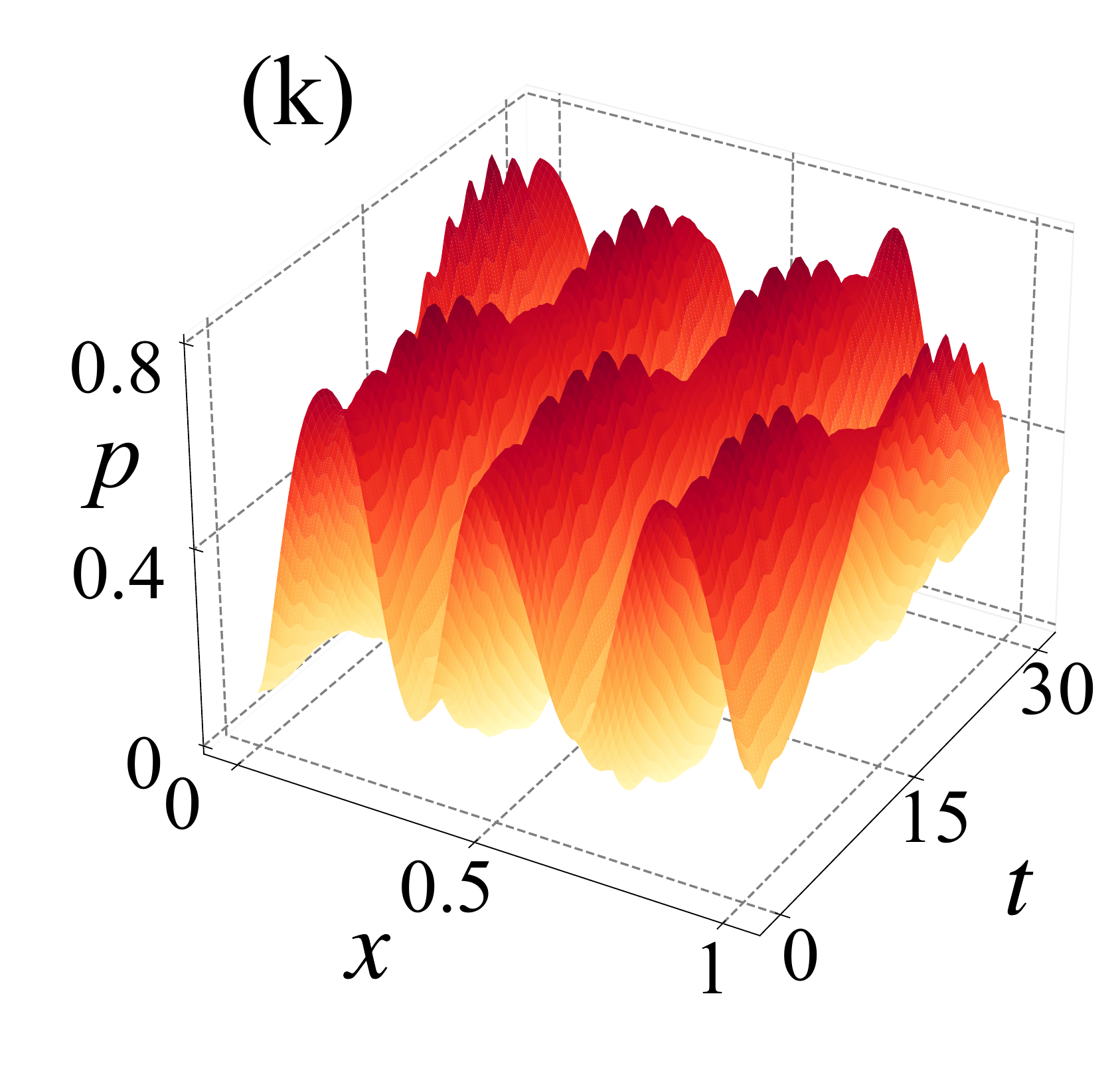}
\includegraphics[scale=0.24]{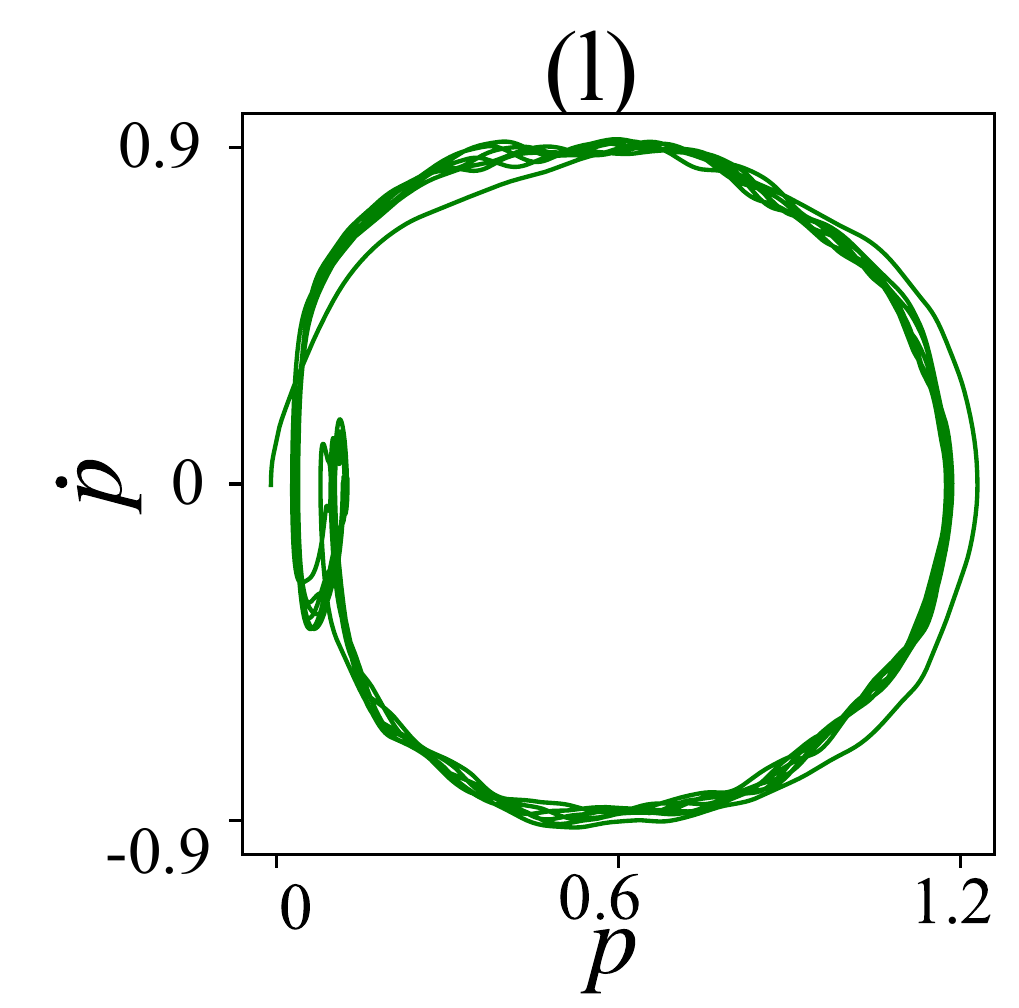}}
\caption{Spatiotemporal evolution of (a,g) the magnetization $m(x,t)$, (c,i) the elastic displacement $u(x,t)$, and (e,k) the polarization $p(x,t)$ for $B_{1}=0.01$ (left panel) and $B_{1}=1$ (right panel). The corresponding phase portraits are shown in (b,h), (d,j), and (f,l), respectively.}
\label{fig1}
\end{figure*}

\begin{figure*}[t]
\centerline
\centerline{ 
\includegraphics[scale=0.255]{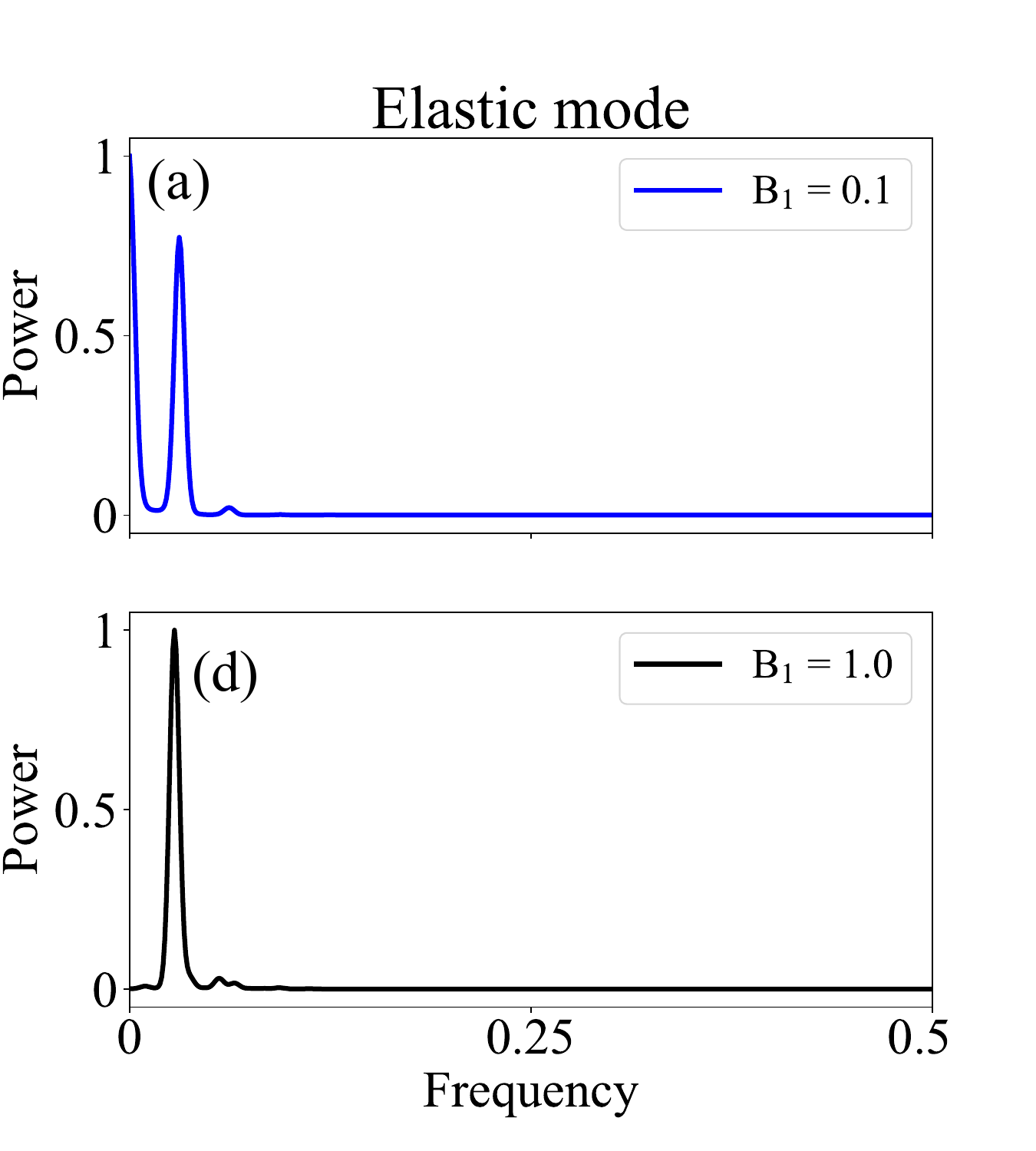}
\includegraphics[scale=0.255]{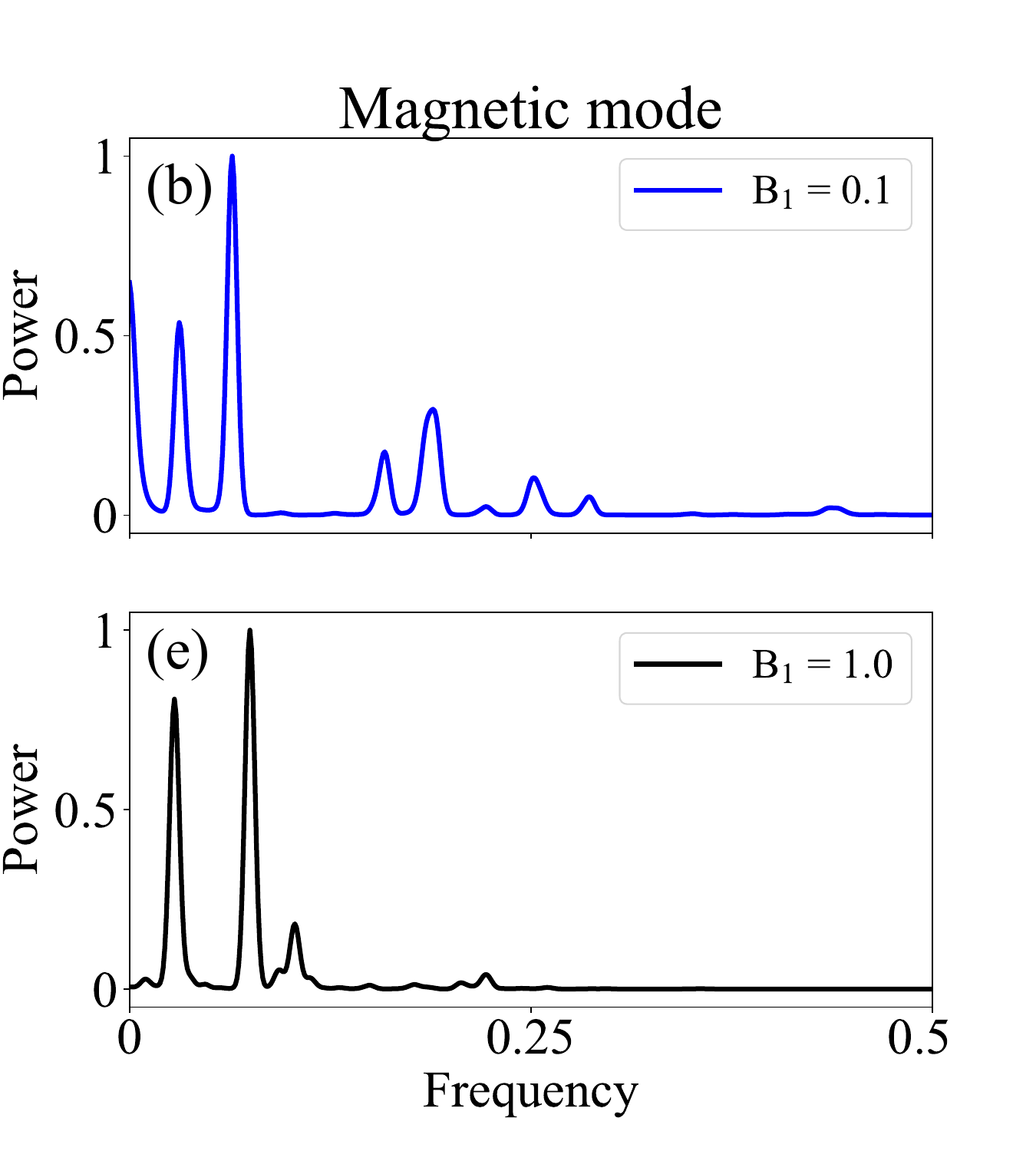}
\includegraphics[scale=0.255]{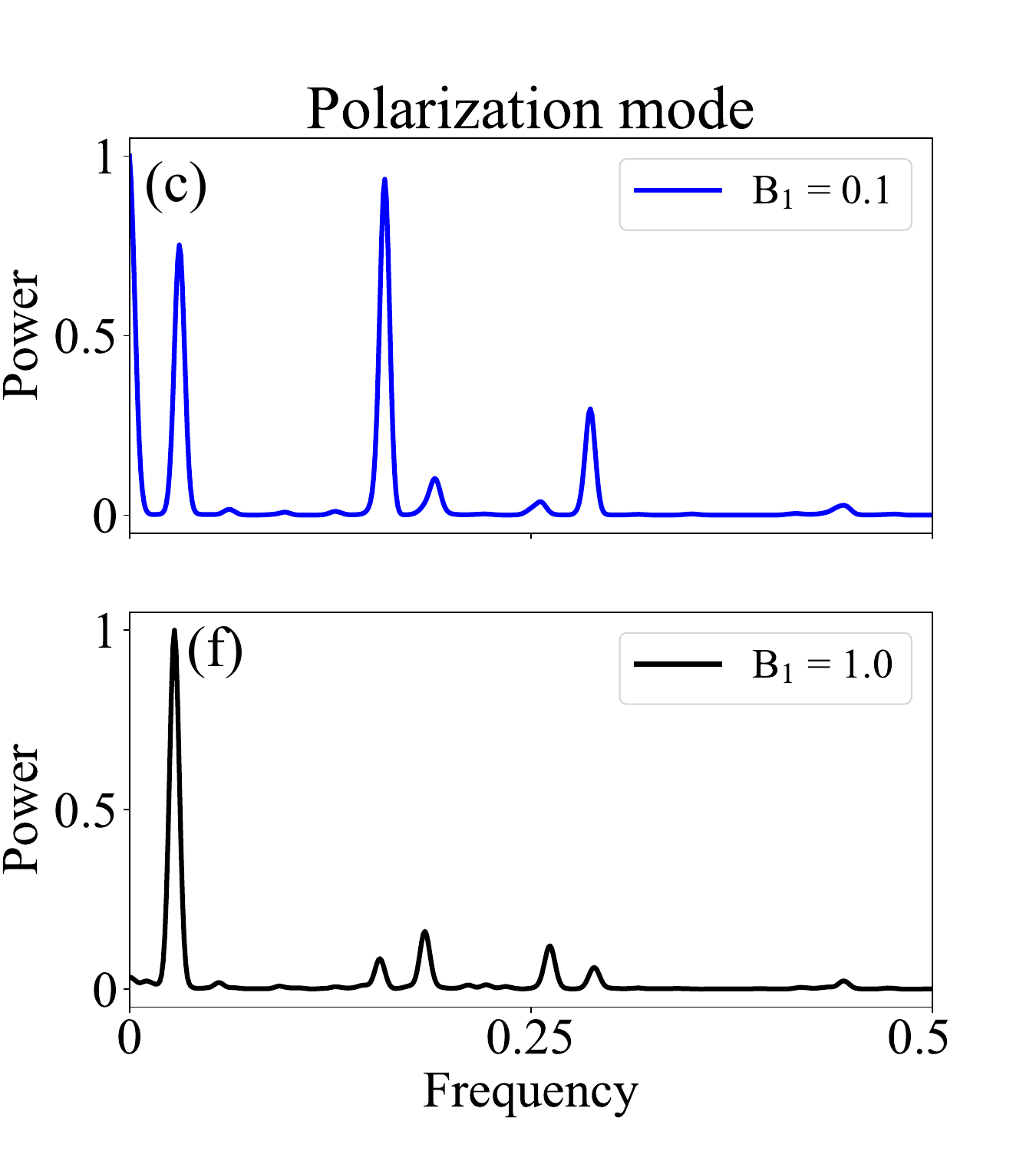}}
\caption{Power spectra of the elastic displacement, magnetization, and polarization modes. for  and strong magnetoelastic coupling. The top row shows the frequency-domain response in the weak ($B_1=0.1$) while the bottom row corresponds to the strong ($B_1=1.0$) magnetoelastic coupling regime.
}
\label{fig2}
\end{figure*}

Fig.~\ref{fig1} illustrates the spatiotemporal evolution and phase space structure of the coupled magnetoelastic system as the magnetoelastic coupling strength $B_1$ is increased from the weak coupling regime ($B_1=0.01$) to the strong coupling regime ($B_1=1$), revealing a systematic transition from weakly nonlinear to strongly nonlinear dynamics across all coupled degrees of freedom. In the weak coupling limit, the magnetization $m(x,t)$ exhibits smooth, low amplitude, and spatially coherent oscillations as seen from Fig.~\ref{fig1}(a). It characterizes linear or weakly nonlinear spin wave dynamics dominated by exchange and anisotropy energies, with only weak strain induced modulation. The corresponding magnetic phase portrait in Fig.~\ref{fig1}(b) consists of closed invariant loops, indicating quasiperiodic motion on regular tori and the absence of strong nonlinear mode coupling. By contrast, the elastic displacement $u(x,t)$ and polarization $p(x,t)$ already display appreciable nonlinear modulation even in the weak coupling regime as seen from Figs.~\ref{fig1}(c) and \ref{fig1}(e) respectively. These results reflect the intrinsic anharmonicity of the elastic and polarization modes. The corresponding phase portraits in Figs.~\ref{fig1}(d) and \ref{fig1}(f) show strongly deformed but closed trajectories, confirming regular yet nonlinear oscillatory dynamics. Upon increasing the coupling to $B_1=1$, the magnetization vector develops pronounced amplitude modulation and spatial inhomogeneity as seen from Fig.~\ref{fig1}(g), indicating strong nonlinear mode coupling induced by the magnetoelastic interaction as the term $B_1 m^2$ becomes comparable to the intrinsic magnetic energy scales. The corresponding magnetic phase portrait in Fig.~\ref{fig1}(h) remains closed but becomes strongly distorted, signaling the emergence of a stable, strongly anharmonic limit cycle rather than chaotic motion. Similarly, both the elastic displacement and polarization exhibit enhanced spatiotemporal modulation and spatial disorder for $B_1 = 1$  as observed from Figs.~\ref{fig1}(i) and \ref{fig1}(k). These results reflect the cooperative nonlinear feedback between magnetic, elastic, and ferroelectric subsystems mediated by magnetoelastic and strain induced magnetoelectric couplings. The corresponding phase portraits in Figs.~\ref{fig1}(j) and \ref{fig1}(l), remain bounded and closed, indicating a crossover from weakly nonlinear quasiperiodic dynamics to strongly nonlinear, yet regular, limit cycle like  behaviour. Overall, Fig.~\ref{fig1} demonstrates that magnetoelastic coupling acts as an efficient control parameter for tuning the degree of nonlinearity and spatiotemporal complexity of collective excitations in multiferroic media, enabling controlled access to strongly anharmonic wave dynamics without inducing chaotic behavior in the present parameter regime.

Fig.~\ref{fig2} presents the power spectra of the elastic displacement, magnetization, and polarization modes in the weak and strong coupling regimes, providing the frequency-domain counterpart of the spatiotemporal dynamics shown in Fig.~\ref{fig1}. For $B_1=0.1$, the spectra of all three modes shown in Figs.~\ref{fig2}(a) –~\ref{fig2}(c) are dominated by a small number of sharp, well separated peaks, indicating nearly monochromatic or weakly multimode oscillations. This behavior is consistent with the smooth, quasiperiodic spatiotemporal patterns and closed invariant loops observed in Figs.~\ref{fig1}(a) –~\ref{fig1}(f). Moreover, these results reflect weak nonlinear mode coupling and the predominance of the intrinsic eigen frequencies of the uncoupled magnetic, elastic, and ferroelectric subsystems. Upon increasing the coupling to $B_1=1.0$, the spectra broaden with some minor peaks are observed in Figs.~\ref{fig2}(d) –~\ref{fig2}(f) for all three modes. These indicate enhanced nonlinear mode coupling and energy transfer between magnetic, elastic, and ferroelectric excitations mediated by magnetoelastic and strain-induced magnetoelectric interactions. Moreover, the appearance of multiple harmonics and combination frequencies reflects strong anharmonicity and intermode frequency mixing, consistent with the distorted limit cycle like behaviour  and pronounced amplitude modulation as observed in Figs.~\ref{fig1}(g) –~\ref{fig1}(l). Importantly, despite the spectral broadening, the spectra remain composed of discrete peaks rather than a continuous broadband background, indicating that the dynamics remains regular and strongly nonlinear rather than chaotic, in the parameter regime considered here. Collectively, Fig.~\ref{fig1} and Fig.~\ref{fig2} demonstrate that increasing magnetoelastic coupling drives a coherent transition from weakly nonlinear, nearly monochromatic oscillations to strongly anharmonic, multimode dynamics, while preserving phase coherence across magnetic, elastic, and ferroelectric degrees of freedom. The correlated spectral evolution across all three subsystems further confirms that magnetoelastic coupling provides an efficient pathway for coherent energy redistribution among spin, lattice, and polarization degrees of freedom.

\section{Magnetoelastic Dispersion Relation}
Understanding the dispersion characteristics of coupled magnetoelastic and electromagnetic excitations is central to describe the dynamical behavior of hexagonal multiferroic materials, where magnetic, elastic, and electric degrees of freedom interact on comparable energy scales. In such systems, static and dynamic magnetoelectric couplings mediate the interconversion between lattice deformations, spin precession, and polarization oscillations thereby giving rise to hybrid quasiparticles such as magnon–phonon and magnon–polariton modes. These hybrid excitations underpin a wide range of functionalities, including strain-controlled spin dynamics, electrically tunable magnon propagation, and the emergence of strongly nonlinear magnetoelastic solitons.  

Assuming plane-wave solutions of the form $f(x,t)=f_0 e^{i(kx-\omega t)}$, we substitute $\partial_t\!\to\!-i\omega$ and $\partial_x\!\to\!ik$ into Eqs.~(\ref{eq15})--(\ref{eq21}) and linearize about the uniform equilibrium magnetization $\mathbf{m}_0 = M_s \hat{z}$, yielding
\begin{align}
\label{eq22}
-i\omega m_x &= \gamma M_s \left( \mathcal{A}k^2 m_y + \gamma_2 p_y \right), \\
\label{eq23}
-i\omega m_y &= -\gamma M_s \left( \mathcal{A}k^2 m_x + \gamma_2 p_x \right).
\end{align}
The elastic equations reduce to
\begin{align}
\label{eq24}
-\rho \omega^2 u_x &= -k^2 \mathcal{C}_{11} u_x - k^2 \mathcal{B}_1 m_x - k^2 \gamma_1 p_x, \\
\label{eq25}
-\rho \omega^2 u_y &= -k^2 \mathcal{C}_{66} u_y - k^2 \mathcal{B}_1 m_y - k^2 \gamma_1' p_y, \\
\label{eq26}
-\rho \omega^2 u_z &= -k^2 \mathcal{C}_{44} u_z - k^2 \mathcal{B}_4 m_z - k^2 \gamma_1 p_z .
\end{align}
The polarization dynamics become
\begin{align}
\label{eq27}
-i\omega p_x &= -\Gamma(\kappa + \delta k^2)p_x  + i\Gamma \gamma_1 k\, u_x + \Gamma \gamma_2 m_x, \\
\label{eq28}
-i\omega p_y &= -\Gamma(\kappa + \delta k^2)p_y + i\Gamma \gamma_1' k\, u_y + \Gamma \gamma_2 m_y .
\end{align}

\noindent
Here $\delta$ controls the gradient-induced polarization stiffness, while $\gamma_1$ and $\gamma_1'$ parameterize the anisotropic electroelastic coupling allowed by $6mm$ symmetry. Collecting Eqs.~(\ref{eq22})--(\ref{eq28}), the coupled system can be written compactly as
$\mathcal{M}(k,\omega)\Psi = 0$, with state vector $\Psi = (m_x, m_y, u_x, u_y, u_z, p_x, p_y)^{T}$. The corresponding dynamical matrix $\mathcal{M}(k,\omega)$ can be written as

\begin{widetext}
\setlength{\arraycolsep}{0.4pt}
\renewcommand{\arraystretch}{2}
\begin{equation}
\begin{small}
\label{eq30}
\mathcal{M}(k,\omega) =
\begin{pmatrix}
-i\omega & \gamma M_s \mathcal{A}k^2 & 0 & 0 & 0 & 0 & \gamma M_s \gamma_2 \\
\gamma M_s \mathcal{A}k^2 & -i\omega & 0 & 0 & 0 & \gamma M_s \gamma_2 & 0 \\
-k^2 B_1 & 0 & -\rho \omega^2 - k^2 \mathcal{C}_{11} & 0 & 0 & -k^2 \gamma_1 & 0 \\
0 & -k^2 B_1 & 0 & -\rho \omega^2 - k^2 \mathcal{C}_{66} & 0 & 0 & -k^2 \gamma_1' \\
-k^2 B_4 & 0 & 0 & 0 & -\rho \omega^2 - k^2 \mathcal{C}_{44} & 0 & 0 \\
\Gamma \gamma_2 & 0 & i\Gamma \gamma_1 k & 0 & 0 
    & -i\omega + \Gamma(\kappa + \delta k^2) & 0 \\
0 & \Gamma \gamma_2 & 0 & i\Gamma \gamma_1' k & 0 
    & 0 & -i\omega + \Gamma(\kappa + \delta k^2)
\end{pmatrix}
\end{small}
\end{equation}
\end{widetext}
The dispersion relations are obtained from the secular equation
\begin{equation}
\det \mathcal{M}(k,\omega) = 0.
\end{equation}

\begin{figure*}[t]
\centerline
\centerline{ 
\includegraphics[scale=0.27]{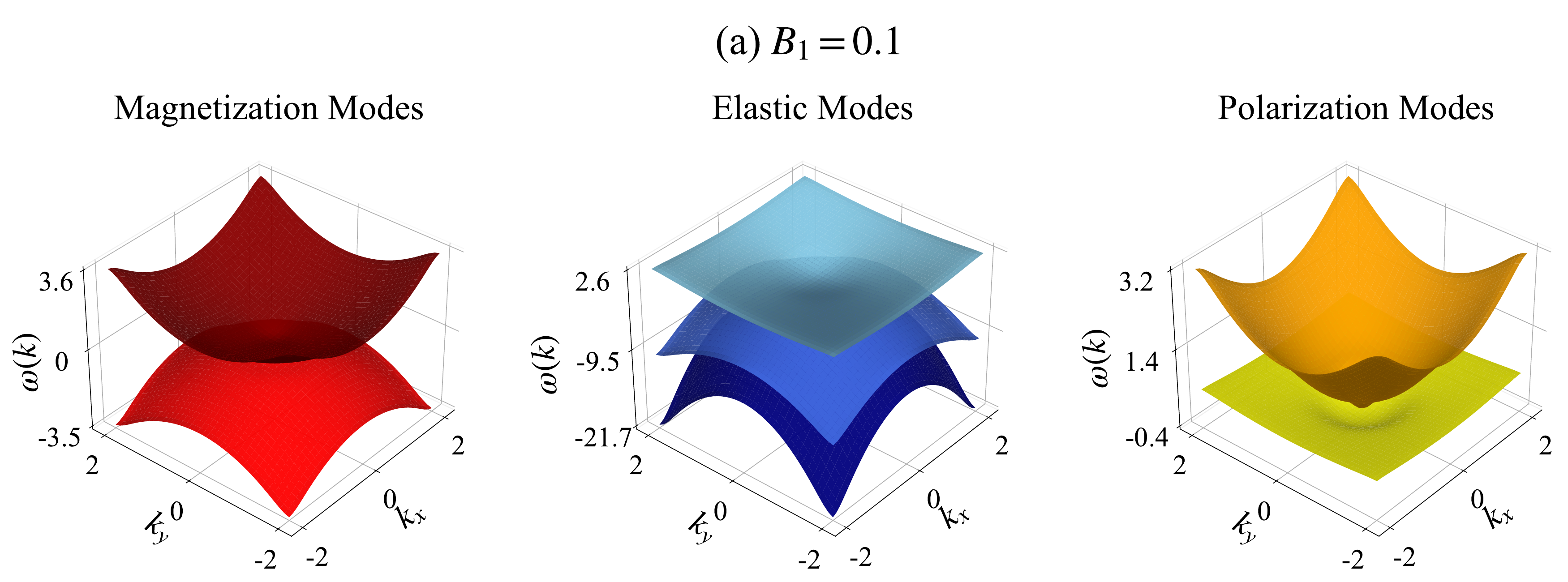}
\includegraphics[scale=0.27]{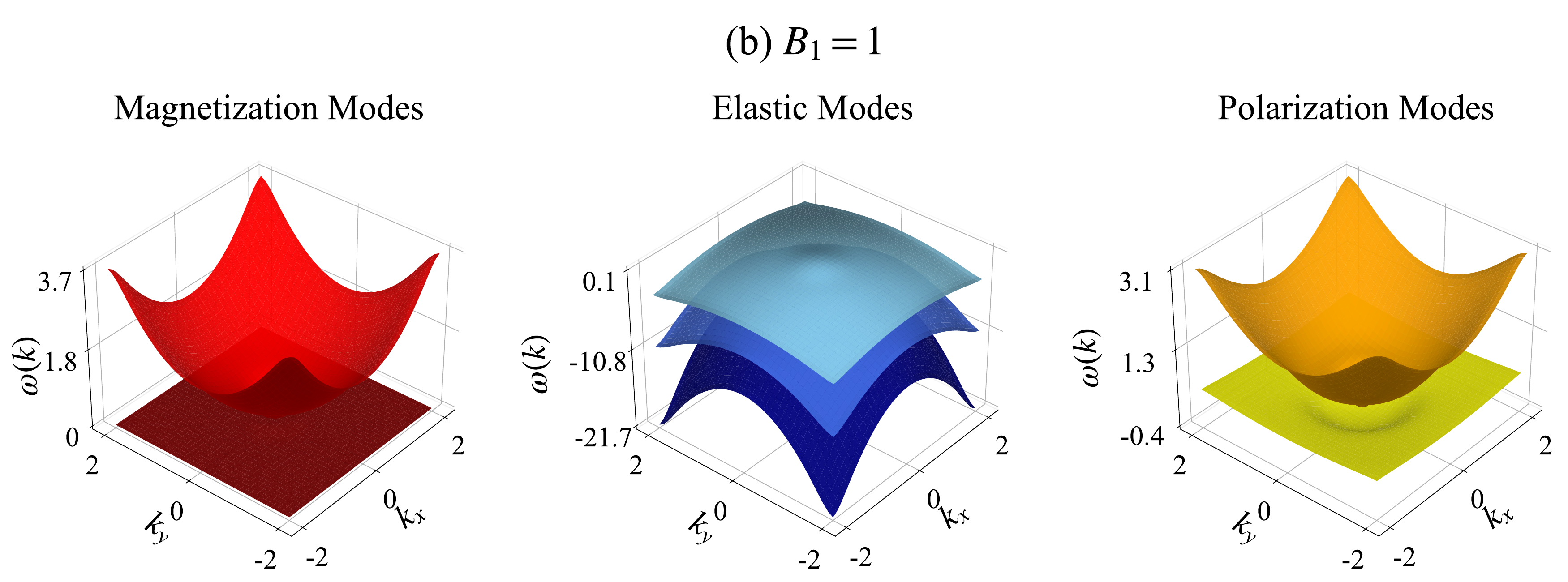}
\includegraphics[scale=0.27]{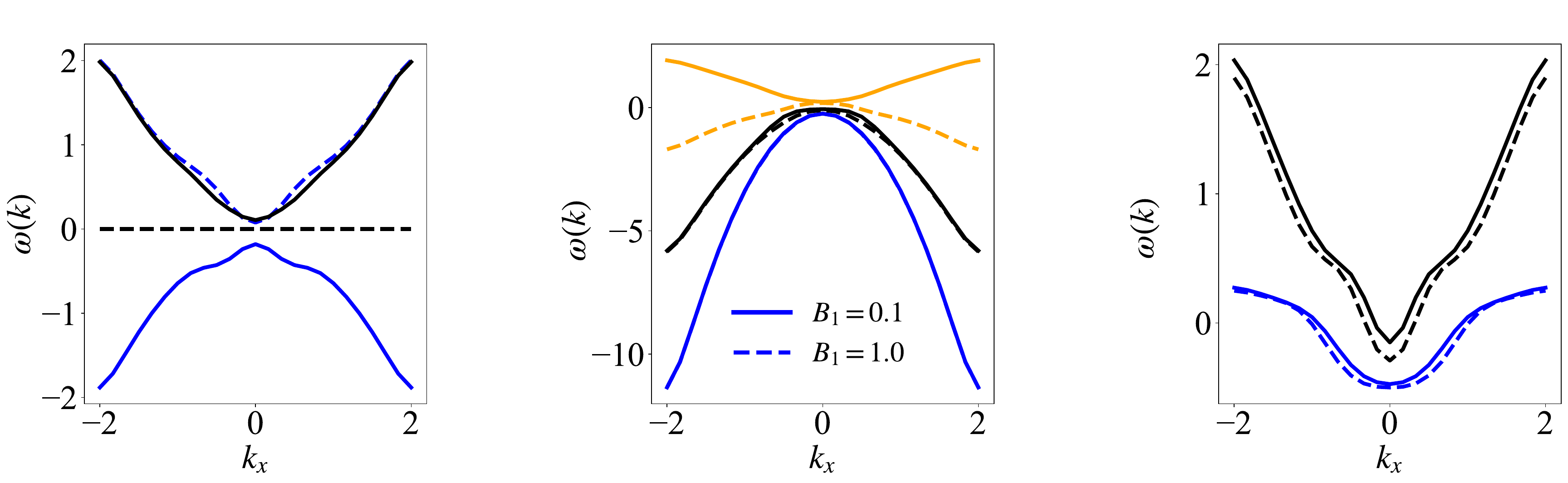}
}
\caption{Dispersion relations of the coupled magnetoelastic system. The top and middle rows show the three-dimensional band surfaces $\omega(\mathbf{k})$ of the magnetic, elastic, and ferroelectric modes in the Brillouin zone for different magnetoelastic coupling strengths $B_1$. The bottom row shows one-dimensional cuts along $k_x$ for $B_1=0.1$ (solid line) and $B_1=1.0$ (dashed line). }
\label{fig3}
\end{figure*}

The hybrid eigenmodes separate into predominantly magnetic, elastic, and ferroelectric branches in the weak-coupling limit, but hybridize at finite $\mathcal{B}_{1,4}$ and $\gamma_2$. The low-frequency sector contains gapless acoustic phonons with linear dispersion, renormalized by magnetoelastic interactions. The polarization modes are gapped and acquire mixed electroelastic character due to the relaxational and gradient terms, while exchange-dominated magnons ($\propto \mathcal{A}k^2$) hybridize with both phonons and polarization modes, producing avoided crossings and magnon–phonon–polariton branches. In addition, crystal symmetry further constrains the excitation spectrum. The $6mm$ symmetry enforces anisotropic elastic constants ($\mathcal{C}_{11}\neq \mathcal{C}_{66}$) while protecting the degeneracy of the in-plane magnetic components $(m_x,m_y)$ and transverse acoustic phonons in the absence of symmetry breaking couplings. Finite magnetoelastic coupling $B_1$ and anisotropic electroelastic interactions $\gamma_1\neq\gamma_1'$ lift these degeneracies, leading to direction dependent mode splitting and anisotropic hybridization~\cite{Nye1985}. The combined action of magnetic, elastic, and polarization couplings therefore determines the topology of the hybrid excitation spectrum and enables tunable magnon–phonon–polariton transport in hexagonal multiferroics.

Fig.~\ref{fig3} illustrates these dispersion characteristics. For $B_1=0.1$, the magnetic, elastic, and polarization branches remain well separated and exhibit nearly parabolic dispersions, reflecting the dominant roles of exchange interactions, elastic stiffness, and ferroelectric restoring forces. However, $B_1=1.0$ produces pronounced renormalization of the dispersion surfaces. In particular, the magnetic and elastic branches show curvature modifications and avoid crossings, indicating strong magnon–phonon hybridization mediated by magnetostrictive interactions, while the polarization modes are renormalized indirectly through strain mediated magnetoelectric coupling. One dimensional cuts along $k_x$ further emphasize the coupling-induced shifts and softening of the excitation branches. The enhanced mode mixing at strong coupling is consistent with the multiple spectral peaks observed in Fig.~\ref{fig2}, reflecting energy exchange among magnetic, elastic, and polarization degrees of freedom. On a collective note, these interaction driven modifications of the dispersion relations provide a spectral basis for the broadband power spectra and complex spatiotemporal dynamics observed at large $B_1$, underscoring magnetoelastic coupling as the primary mechanism governing synchronized nonlinear behavior across the coupled subsystems.

\section{Magnetoelastic Soliton Excitations}
The strongly nonlinear behavior observed in the coupled magnetoelastic–ferroelectric dynamics motivates the investigation of coherent nonlinear excitations supported by the same microscopic interactions. In multiferroic media, where exchange driven spin waves, elastic phonons, and polarization modes coexist and hybridize, the competition between dispersion and nonlinearity is expected to give rise to soliton excitations. Such solitons represent dynamically stable, particle like objects that propagate without dispersion and may be viewed as nonlinear counterparts of the hybrid magnon–phonon–polariton modes discussed in Sec.~III. Understanding their emergence is not only of fundamental interest but also due to its potential applications in reconfigurable spintronic, straintronic, and magnetoelectric signal processing, where information carriers immune to dispersion and moderate disorder are highly desirable.

In this section, we develop a unified theoretical framework for magnetoelastic soliton formation in hexagonal multiferroic systems. Building directly on the coupled equations of motion derived in Sec.~I and the dispersion analysis of Sec.~III, we demonstrate how strong magnetoelastic and magnetoelectric couplings transform the linear hybrid modes into nonlinear solitary waves.  To isolate nonlinear wave excitations, we consider a quasi one-dimensional geometry corresponding to propagation along the hexagonal $c$ axis i.e. along $z$ direction, relevant for multiferroic nanowires and channelized domain-wall modes. We take
\begin{align}
\nonumber
\mathbf{u}(z,t) &= u(z,t)\hat{z}, \\ \nonumber \quad
\mathbf{m}(z,t) &= m_x(z,t)\hat{x} + m_z(z,t)\hat{z}, \\ \quad
\mathbf{p}(z,t) &= p_z(z,t)\hat{z}.
\label{eq31}
\end{align}
In this limit, the dominant strain component is $\epsilon_{zz}=\partial_z u$, and the elastic and magnetoelastic energy densities reduce to $ 
\mathcal{H}_{\mathrm{elas}}  \approx \frac{\mathcal{C}_{33}}{2}(\partial_z u)^2$ and $
\mathcal{H}_{\mathrm{mag\text{-}elas}} \approx -B_2 m_z^2 \partial_z u
$. The polarization field can be eliminated algebraically using the constitutive relation obtained from minimizing $\mathcal{H}_{\mathrm{elec}}+\mathcal{H}_{\mathrm{me}}$,
\begin{equation}
p_z = \chi_e (E_z - \gamma_2 m_z),
\label{eq32}
\end{equation}
which explicitly demonstrates how electric fields modulate magnetic dynamics through magnetoelectric coupling.

Under these assumptions, the coupled equations of motion, Eqs.~(\ref{eq15})-(\ref{eq21}), reduce to
\begin{align}
\rho \ddot{u} &= \mathcal{C}_{33}\partial_z^2 u - B_2 \partial_z (m_z^2),
\label{eq33}\\
\tau\ddot{m}_z &= \mathcal{A} \partial_z^2 m_z
- \kappa_0 m_z
- 2B_2 m_z \partial_z u
+ \eta E_z .
\label{eq34}
\end{align}

\noindent where, $\tau = \frac{1}{\chi_m}$, $\eta = \gamma_2 \chi_e$ and $\kappa_0 = \left(\mathcal{K}_u + \gamma_2^2\chi_e\right)$. For coherent traveling excitations, we introduce the comoving coordinate $\xi=z-vt$ and assume $u(z,t)=U(\xi)$ and $m_z(z,t)=M(\xi)$. In the comoving frame, the derivatives transform as, $\partial_t^2 u = v^2 U''$, $\partial_z^2 u = U''$. Now integrating Eq.~(\ref{eq33}), we get
\begin{equation}
U(\xi)=\frac{B_2}{\mathcal{C}_{33}-\rho v^2}M^2(\xi),
\label{eq35}
\end{equation}
Substituting this into Eq.~(\ref{eq34}) leads to a closed nonlinear equation for the magnetization amplitude,
\begin{equation}
\alpha M''(\xi) - \left\{\kappa_0 + \lambda M^2(\xi)\right\}M(\xi) + \eta E_z = 0,
\label{eq36}
\end{equation}
where, $\alpha=v^2\tau-\mathcal{A}$ and $
\beta\equiv\lambda=\frac{2B_2^2}{\mathcal{C}_{33}-\rho v^2}.$
Eq.~(\ref{eq36}) is a driven Duffing-type nonlinear oscillator equation in spatial form, with the cubic nonlinearity originating entirely from magnetoelastic coupling. This result establishes a direct link between the microscopic magnetostriction terms in Sec.~II and the emergence of nonlinear solitary waves. 
\begin{figure*}[t]
\centerline
\centerline{ 
\hspace{-0.5mm}
\includegraphics[scale=0.42]{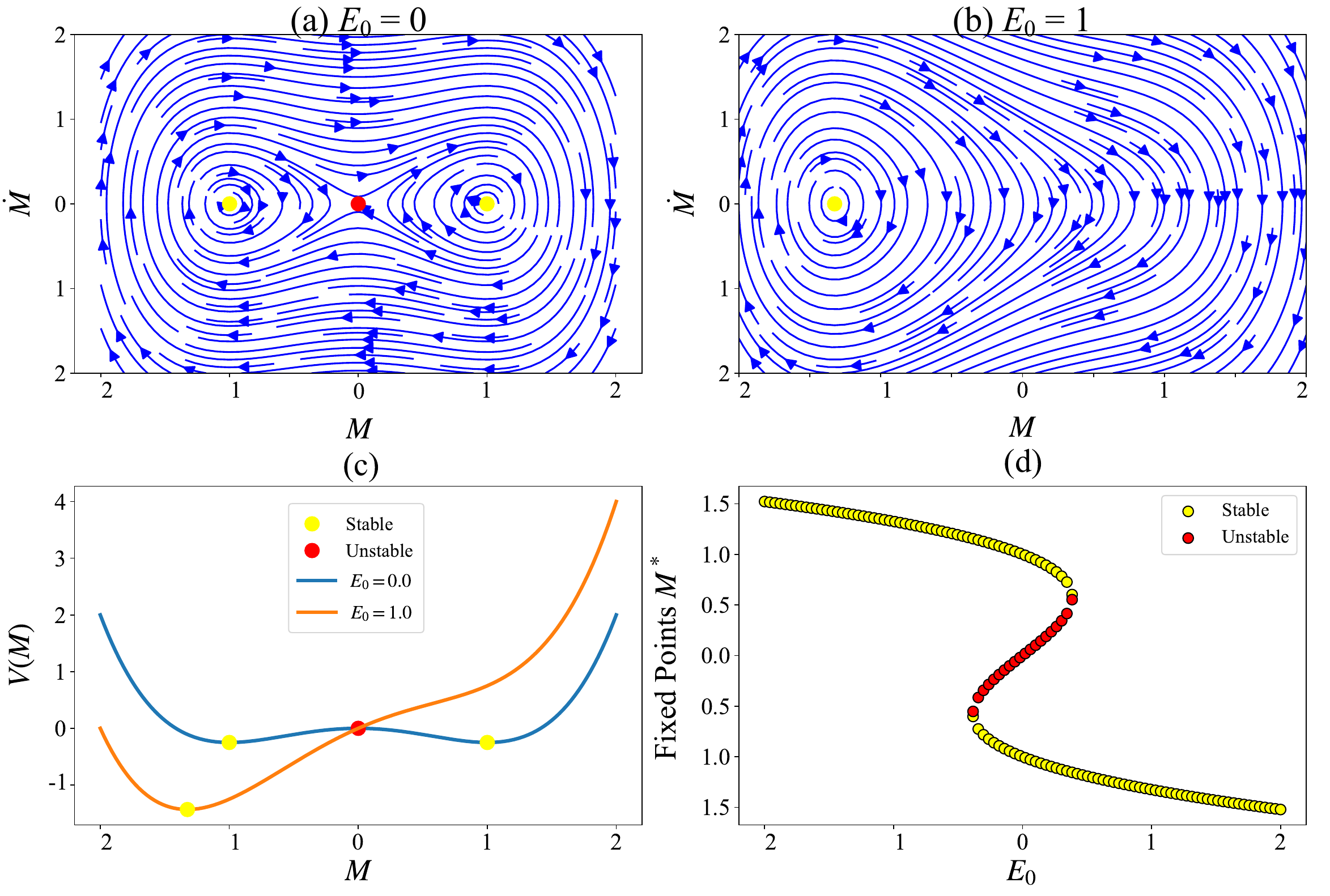}
}
\caption{
Streamlines of the phase portrait in the $(M,\dot{M})$ plane for (a) $E_0=0$ and (b) $E_0=1$. 
Stable fixed points are indicated by yellow circles, and the unstable saddle point by red circles. 
(c) Variation of effective potential $V(M)$ with $M$ for $E_0=0$ (blue) and $E_0=1$ (orange). 
(d) Bifurcation diagram of the fixed points $M^\ast$ as a function of $E_0$, showing the field-driven annihilation and creation of stable (yellow) and unstable (red) solutions.
}
\label{fig4}
\end{figure*}

\subsection{Soliton solutions in absence of $E_z$}
We first set $E_z = 0$. In view of this Eq.~\eqref{eq36} reduces to
\begin{equation}
    \alpha M''(\xi) - \big\{\kappa _0+ \lambda M^2(\xi)\big\}M(\xi) = 0.
    \label{eq37}
\end{equation}
Multiplying Eq.~\eqref{eq37} by $M'(\xi)$ and integrating once yields the energy like integral
\begin{equation}
    \frac{\alpha}{2}\big\{M'(\xi)\big\}^2 - \frac{\kappa_0}{2}M^2(\xi) - \frac{\lambda}{4}M^4(\xi) = C,
    \label{eq38}
\end{equation}
where $C$ is an integration constant. For localized soliton solutions with $M(\xi)\to 0$ and $M'(\xi)\to 0$ as $|\xi|\to\infty$, we set $C=0$. Thus,
\begin{equation}
    M'(\xi) = M(\xi)\sqrt{\left[  \frac{\lambda}{2\alpha}M^2(\xi)+\frac{\kappa_0}{\alpha}\right]}.
    \label{eq39}
\end{equation}
We consider the standard bright soliton ansatz
$M(\xi)=M_0\,\mathrm{sech}(\kappa\xi),$
with
\begin{equation}
\kappa=\sqrt{\frac{\kappa_0}{\alpha}} \qquad \text{and} \quad
M_0=\sqrt{\frac{-2\kappa_0}{\lambda}},
\label{eq40}
\end{equation}
provided $\kappa_0<0$ and $\lambda>0$. These conditions correspond physically to a softening of the magnetic anisotropy due to magnetoelectric renormalization and sufficiently strong magnetoelastic coupling. The soliton width is given by $ w = \frac{1}{\kappa} = \sqrt{\frac{\alpha}{|\kappa_0|}}$. 

\subsection{Soliton solutions in presence of $E_z$}
For a system with $E_z\neq0$, the soliton profile persists with an electric-field-dependent amplitude determined by
\begin{equation}
\kappa_0 M + \lambda M^3 = \eta E_z,
\label{eq41}
\end{equation}
The asymptotic magnetization $M_\infty$, determined by Eq.~(\ref{eq41}), governs the existence and stability of localized solutions. This relation explicitly demonstrates the electric field control of soliton properties. To analyze its solution structure, we rewrite Eq.~(\ref{eq41}) in the cubic form by dividing through by $\lambda \neq 0$,
\begin{equation}
M^3 + p M + q = 0 ,
\label{eq42}
\end{equation}
with coefficients $
p = \frac{\kappa_0}{\lambda}$ and $
q = -\frac{\eta E_z}{\lambda}$. The discriminant of this equation is given by
$\Delta = -4 p^3 - 27 q^2$. Substituting the value of p and q and multiplying by $\lambda^3$, the discriminant can be written compactly as
\begin{equation}
\Delta = -4 \kappa_0^3 \lambda
         - 27 \eta^2 E_z^2 \lambda^2 .
\label{eq43}
\end{equation}
The number of real solutions is therefore characterized  in three different conditions as mentioned in the Table~\ref{tab1}
\begin{table}[t]
\centering
\caption{Classification of equilibrium solutions of the cubic equation in terms of the discriminant $\Delta$, indicating the corresponding stability regime and underlying bifurcation structure.}
\begin{tabular}{c c c}
\hline\hline
Discriminant $\Delta$ & Number of real roots & Physical regime \\ 
\hline
$\Delta > 0$ & Three distinct  & Multistable regime  \\[4pt]
$\Delta = 0$ & Two coincident  & Saddle-node bifurcation \\[4pt]
$\Delta < 0$ & One  & Monostable regime \\ 
\hline\hline
\label{tab1}
\end{tabular}
\end{table}

At the saddle-node bifurcation, $\Delta = 0$, solving for the electric field, we obtain the critical field
\begin{equation}
E_z^{\mathrm{crit}}
= \pm \frac{2}{3\sqrt{3}\eta}
\frac{|\kappa_0|^{3/2}}{ \sqrt{|\lambda|}} .
\label{eq44}
\end{equation}

The soliton width is given by the effective potential well:
\begin{equation}
(M')^2 = \frac{2}{\alpha} \left[ V(M) - V(M_\infty) \right]
\label{eq45}
\end{equation}
Therefore, the width is approximately:
\begin{equation}
    w \sim \left( \int_{M_\infty}^{M_0} \frac{dM}{\sqrt{(2/\alpha)[V(M) - V(M_\infty)]}} \right)
    \label{eq46}
\end{equation}

\begin{figure*}[t]
\centerline
\centerline{ 
\hspace{-0.5mm}
\includegraphics[scale=0.35]{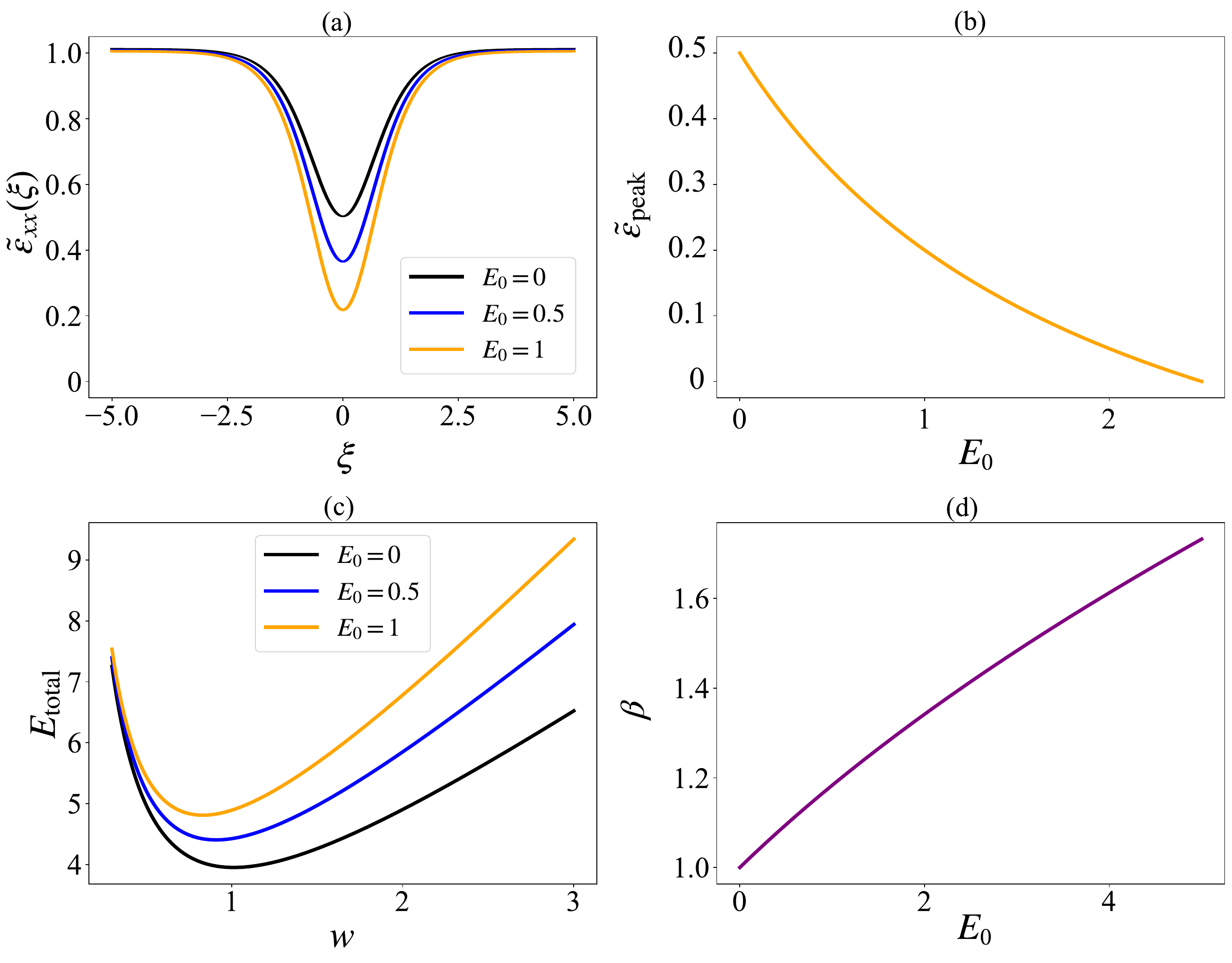}
}
\caption{(a) Spatial profiles of the normalized strain component $\tilde{\varepsilon}_{xx}(\xi)$
associated with the soliton for different electric-field strengths
$E_0=0$, $0.5$, and $1$. (b) Peak strain amplitude $\tilde{\varepsilon}_{\mathrm{peak}}$ as a function
of the electric field $E_0$. (c) Total soliton energy $E_{\mathrm{total}}$ as a function of the soliton width
$w$ for different electric fields and (d) Electric field dependence of the nonlinear coefficient $\beta$.
}
\label{fig5}
\end{figure*}

This integral is field-dependent via  $V(M)$, hence the soliton width is tunable using  $E_z$.
The critical field $E_z^{\mathrm{crit}}$ in Eq.~(\ref{eq44}) gives the boundary between mono and multistable regimes. 
Below this threshold, multiple equilibrium magnetization states coexist, enabling electrically tunable soliton amplitudes. Above the critical field, only a single stable branch survives, suppressing bistability and eliminating localized solutions. Fig.~\ref{fig4} illustrates the phase space structure, effective potential landscape,
and electric field induced bifurcation of the magnetization dynamics obtained by solving Eq.~(\ref{eq36}).
Figs.~\ref{fig4}(a) and \ref{fig4}(b) show the phase portraits in the $(M,\dot{M})$ plane for $E_z=0$ and $E_z=1$ respectively. For $E_z=0$, the system
exhibits a symmetric double-well structure in phase space, corresponding to two
stable fixed points at finite magnetization and an unstable saddle point at the
origin. These fixed points arise from the competition between exchange
stiffness and magnetoelastic nonlinearity and support localized bright soliton
solutions when $\kappa_0<0$ and $\lambda>0$. The separatrix connecting the saddle points defines the homoclinic orbit corresponding to the exact bright-soliton solution $M(\xi) = M_0\,\mathrm{sech}(k\xi)$. When a finite electric field is applied, the phase space symmetry is
broken due to the linear magnetoelectric coupling term proportional to
$E_z$. This tilting of the phase portrait lifts the degeneracy
between the two minima, leading to the annihilation of one stable fixed point
via a saddle-node bifurcation. Consequently, the system undergoes a transition
from a multistable to a monostable regime. These results signify the electric field
control over the magnetization equilibria and the background on which solitons propagate.

The corresponding effective potential $V(M)$ derived from the Eq.~(\ref{eq37}) is shown in Fig.~\ref{fig4}(c). For $E_z=0$, the symmetric double-well potential supports two
energetically equivalent stable equilibria. However, a finite electric field i.e., $E_z = 1$ introduces
an asymmetric bias, lowering one minimum while raising the other, in direct
correspondence with the phase space evolution as already observed in  Figs.~\ref{fig4}~(a) and ~\ref{fig4}(b).
The fixed-point structure as a function of $E_z$ is depicted in Fig.~\ref{fig4}~(d). The bifurcation diagram reveals the characteristic saddle-node
behavior predicted by the discriminant analysis of the cubic equilibrium Eq.~(\ref{eq42}). It is observed that stable and unstable branches
merge at the critical field $E_z^{\mathrm{crit}}$ and beyond that only a single stable equilibrium state is present.
This bifurcation establishes the electric field threshold for magnetization switching and directly determines the existence and stability of magnetoelastic soliton solutions.

Fig.~\ref{fig5} illustrates the electric-field tunability of magnetoelastic soliton
properties arising from the coupled magnetic and elastic degrees of freedom.
Fig.~\ref{fig5}(a) shows the spatial profiles of the normalized longitudinal strain
$\tilde{\varepsilon}_{xx}(\xi)$ associated with the soliton for different values
of the applied electric field $E_0$. For $E_0=0$, the strain profile exhibits a localized depression centered at
$\xi=0$, reflecting the strain generated by the spatially localized magnetic
soliton through magnetoelastic coupling. In presence of electric field, i.e. for  $E_0 = 0.5$ and $1$, the
strain localization becomes more pronounced and its minimum value decreases,
demonstrating that the electric field effectively enhances the magnetoelastic
response by modifying the equilibrium magnetization and elastic displacement.
The corresponding peak strain amplitude
$\tilde{\varepsilon}_{\mathrm{peak}}$ as a function of $E_0$ is presented in Fig.~\ref{fig5}(b). We observe a monotonic
reduction of $\tilde{\varepsilon}_{\mathrm{peak}}$ with increasing electric
field indicates a field-induced renormalization of the effective elastic
stiffness mediated by magnetoelectric coupling. Physically, the applied electric
field shifts the magnetic equilibrium state and reduce the incremental
strain generated by the soliton induced magnetization modulation.

In Fig.~\ref{fig5}(c), the total soliton energy $E_{\mathrm{total}}$ is plotted as a
function of the soliton width $w$ for different values of $E_0$. It is observed that each
curve exhibits a well-defined minimum, corresponding to a stable soliton width
resulting from the balance between exchange-driven dispersion and nonlinear
magnetoelastic interactions. Increasing $E_0$ value shifts the minimum to
larger widths and higher energies, indicating that electric field control allows continuous tuning of both the energetic stability and the characteristic spatial
extent of the soliton. Fig.~\ref{fig5}(d) shows the electric-field dependence of the nonlinear coefficient
$\beta$, which quantifies the strength of the effective cubic nonlinearity in Eq.~(\ref{eq36}). The
monotonic increase of $\beta$ with $E_0$ reflects the enhancement of magnetoelastic nonlinearity due to electric field induced modification of the
magnetization, providing a direct mechanism for electrically controlled
nonlinear spin wave dynamics.

\subsection{Reduction of the magnetization dynamics to nonlinear Schr\"{o}dinger equation (NLSE)}

To describe slowly varying magnetization envelopes and establish a connection with standard soliton theory, we reduce the coupled magnetoelastic equations of motion to an effective nonlinear Schr\"{o}dinger equation (NLSE). This approach enables a systematic analysis of envelope solitons and their tunability through magnetoelastic interactions. We consider wave propagation along the $x$ direction and retain only longitudinal strain,
$\varepsilon_{xx}$ and $
\varepsilon_{yy}\approx \varepsilon_{zz}\approx \varepsilon_{xy}\approx 0 
$. Under this assumption, we can write
\begin{align}
\label{eq47}
\mathcal{H}_{\text{elas}} &\simeq \frac{1}{2}\mathcal{C}_{11}\left(\frac{\partial u_x}{\partial x}\right)^2, \\
\label{eq48}
\mathcal{F}_{\text{mag-elas}} &\simeq (B_1 m_x^2  - B_3 m_z^2) \frac{\partial u_x}{\partial x}.
\end{align}

Thus the elastic equation of motion becomes
\begin{equation}
\rho \frac{\partial^2 u_x}{\partial t^2}
= \frac{\partial }{\partial x}\!\left(\mathcal{C}_{11}\frac{\partial u_x}{\partial x} + B_1 m_x^2 - B_3 m_z^2\right).
\label{eq49}
\end{equation}

The corresponding magnetoelastic effective fields entering the Landau-Lifshitz equation are
\begin{align}
\nonumber
H_{\text{mag-elas},x} &= \frac{2B_1m_x}{M_s}\frac{\partial u_x}{\partial x},\\
\qquad
\nonumber
H_{\text{mag-elas},z} &= -\frac{2B_3m_z}{M_s}\frac{\partial u_x}{\partial x} .
\end{align}

Assuming small transverse fluctuations about the equilibrium magnetization,
$\mathbf{m}\simeq(m_x,m_y,M_s)$ and $  m_x,m_y\ll M_s ,$
the linearized Landau-Lifshitz equations reduce to
\begin{align}
\label{eq50}
\frac{\partial m_x}{\partial t} &= -\gamma M_s \mathcal{A} \frac{\partial^2 m_y}{\partial x^2} \\
\label{eq51}
\frac{\partial m_y}{\partial t} &= \gamma M_s
\left(\mathcal{A} \frac{\partial^2 m_x}{\partial x^2}
+ \frac{2B_1}{M_s}m_x\frac{\partial u_x}{\partial x}\right)
\end{align}

Eliminating $m_y$, the Eq.~(\ref{eq50}) reduces to
\begin{equation}
\frac{\partial^2 m_x}{\partial t^2}
= -\gamma^2 M_s^2 \mathcal{A} ^2 \frac{\partial^4 m_x}{\partial x^4}
- 2\gamma^2 M_s \mathcal{A}  B_1 \frac{\partial^2}{\partial x^2}
\left(m_x\frac{\partial u_x}{\partial x}\right).
\label{eq52}
\end{equation}

In the quasi-static elastic limit, where the phonon dynamics is fast compared to spin dynamics, $\omega_{\mathrm{mag}}\ll\omega_{\mathrm{ph}}$, Eq.~(\ref{eq49}) reduces to
\begin{equation}
\frac{\partial u_x}{\partial x}\simeq -\frac{2B_1}{\mathcal{C}_{11}} m_x\frac{\partial m_x}{\partial x},
\label{eq53}
\end{equation}
Substituting into Eq.~(\ref{eq52}), we obtain the nonlinear wave equation. After eliminating the displacement field in the quasi-static
limit and reducing the Landau–Lifshitz equations, the resulting nonlinear wave equation for $m_x$ can be written as
\begin{equation}
\frac{\partial^2 m_x}{\partial t^2}
+ \gamma^2 M_s^2 \mathcal{A}^2 \frac{\partial^4 m_x}{\partial x^4}
+ \mu\,\frac{\partial^2}{\partial x^2}(m_x^3)=0,
\label{eq54}
\end{equation}
where, the effective nonlinearity factor  $\mu = \frac{4\gamma^2 M_s A B_1^2}{\mathcal{C}_{11}}$.

Considering the solutions of the above equation is of the form 
\begin{equation}
m_x = \epsilon\Psi e^{i(kx-\omega t)}
+ \epsilon\Psi^*e^{-i(kx-\omega t)} + \mathcal{O}(\epsilon^3),
\label{eq55}
\end{equation}
where, $\Psi \equiv \Psi(X,T)$ and $\epsilon \ll 1$ is a small parameter. The variables $(X,T)$ are defined as $X=\epsilon(x-v_g t)$ and $T=\epsilon^2 t$. At leading order $\mathcal{O}(\epsilon)$, we recover the linear spin-wave dispersion and corresponding to the exchange dominated magnon mode, with group velocity respectively are
\begin{align}
\omega^2&=\gamma^2 M_s^2 \mathcal{A}^2 k^4,
\label{eq56}\\
v_g&=\frac{d\omega}{dk}
=2\gamma M_s \mathcal{A} k.
\label{eq57}
\end{align}

At $\mathcal{O}(\epsilon^3)$, removal of secular terms yields
\begin{equation}
i\omega \frac{\partial\Psi}{\partial T}
+\left(v_g^2 + 6 \gamma^2 M_s^2 \mathcal{A}^2 k^2\right)
\frac{\partial^2\Psi}{\partial X^2}
+ 3\mu k^2 |\Psi|^2\Psi=0
\label{eq58}
\end{equation}

Dividing by the carrier frequency $\omega$ to obtain the canonical NLSE form, we arrive at
\begin{equation}
i\frac{\partial\Psi}{\partial T}
+\mathcal{P}\frac{\partial^2\Psi}{\partial X^2}
+\mathcal{Q}|\Psi|^2\Psi=0,
\label{eq59}
\end{equation}
where, the cefficients $\mathcal{P}$ and $\mathcal{Q}$ are defined as 
\begin{align}
\mathcal{P}
\label{eq60}
&=\frac{v_g^2 + 6 \gamma^2 M_s^2 \mathcal{A}^2 k^2}{\omega}\\
\mathcal{Q}
&=\frac{3\mu k^2}{\omega}
=\frac{12\gamma^2 M_s \mathcal{A} B_1^2k^2}{\mathcal{C}_{11}\omega}
\label{eq61}
\end{align}
originates from magnetoelastic interactions mediated by longitudinal lattice distortions. Notably, the magnitude and sign of $\mathcal{Q}$ can be tuned through the magnetoelastic constants $B_1$ and elastic modulus $\mathcal{C}_{11}$, both of which are sensitive to external electric and strain fields in multiferroic media.

\begin{figure*}[t]
\centerline
\centerline{ 
\includegraphics[scale=0.22]{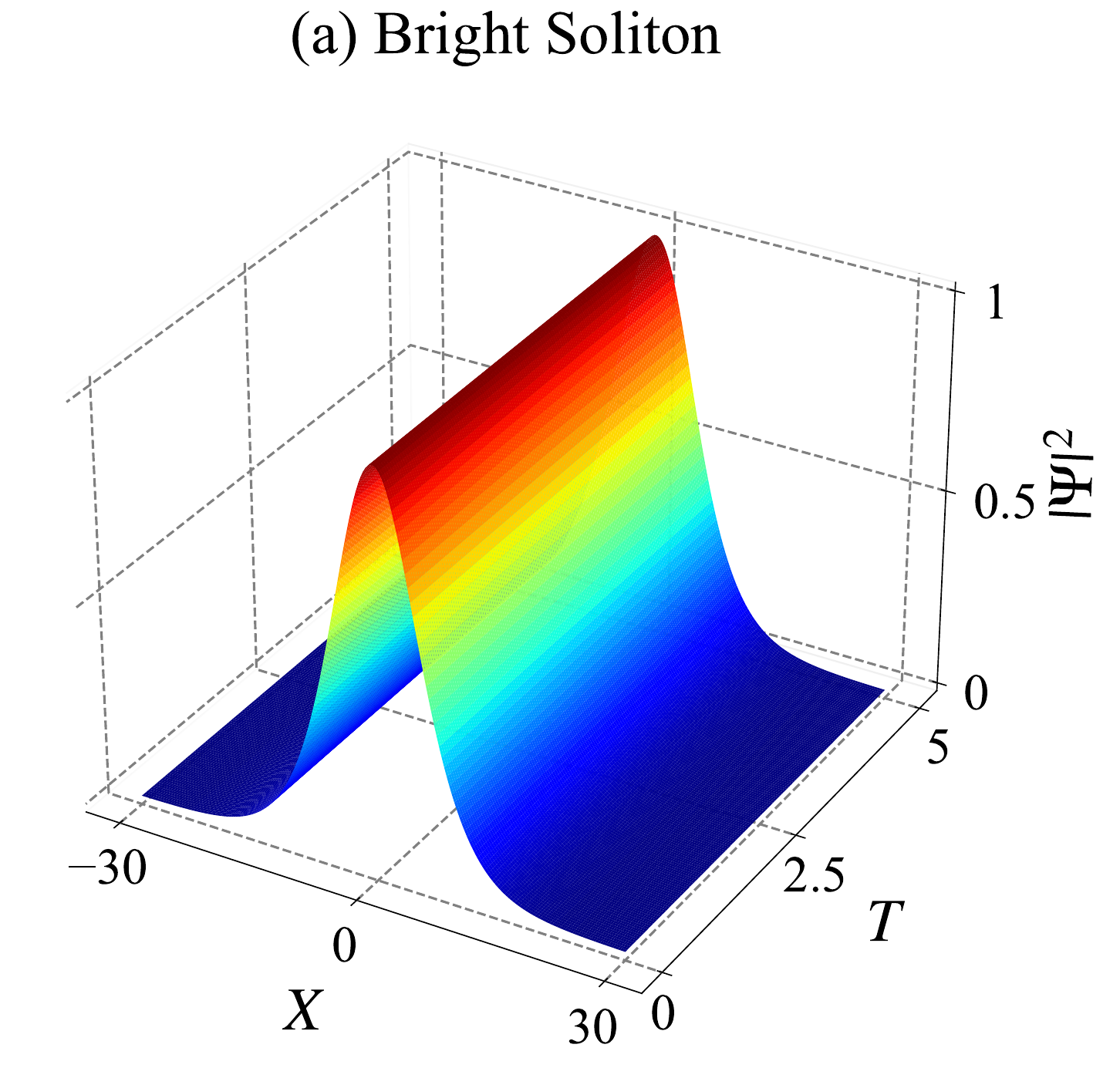}
\includegraphics[scale=0.22]{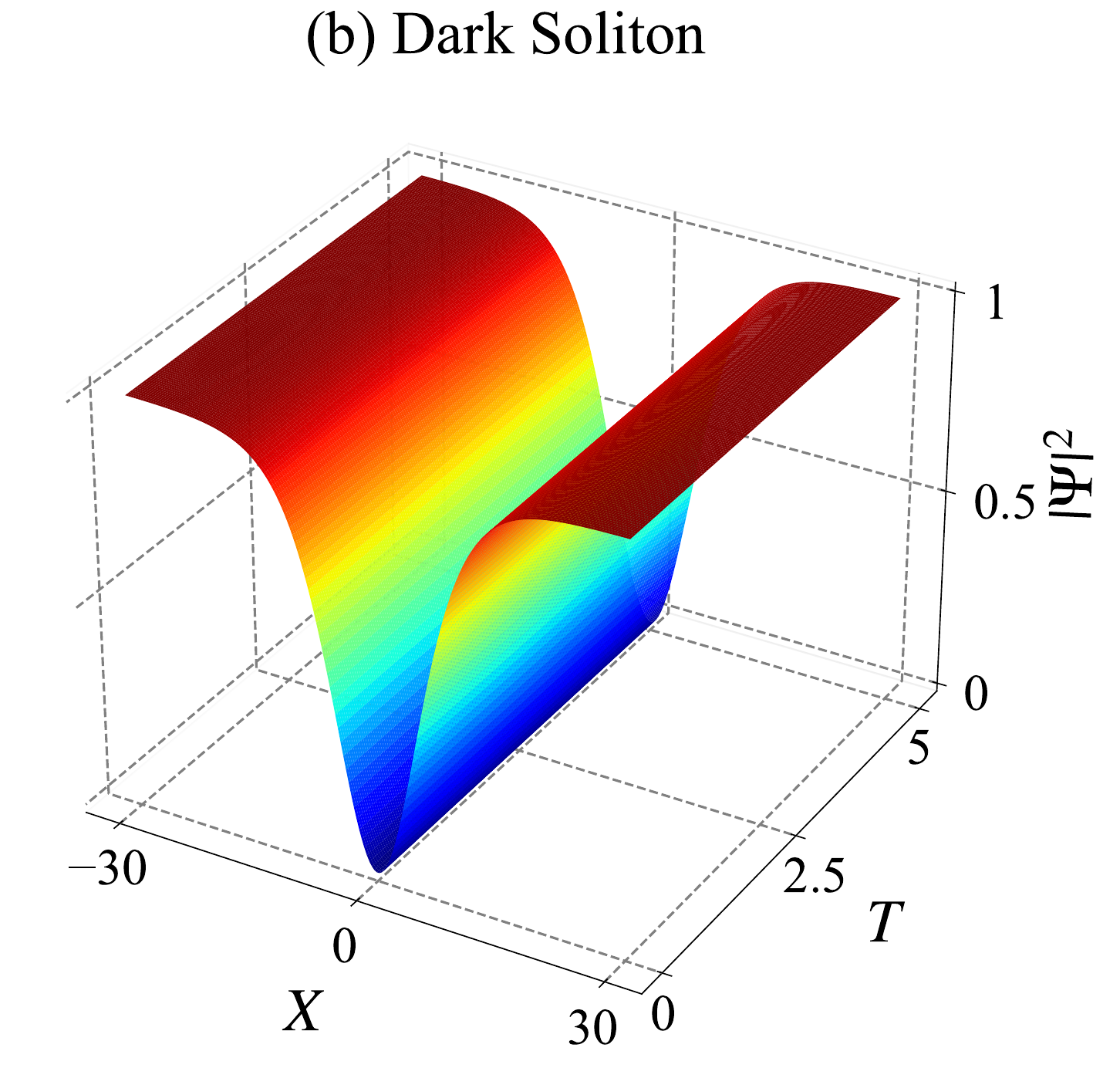}
\includegraphics[scale=0.22]{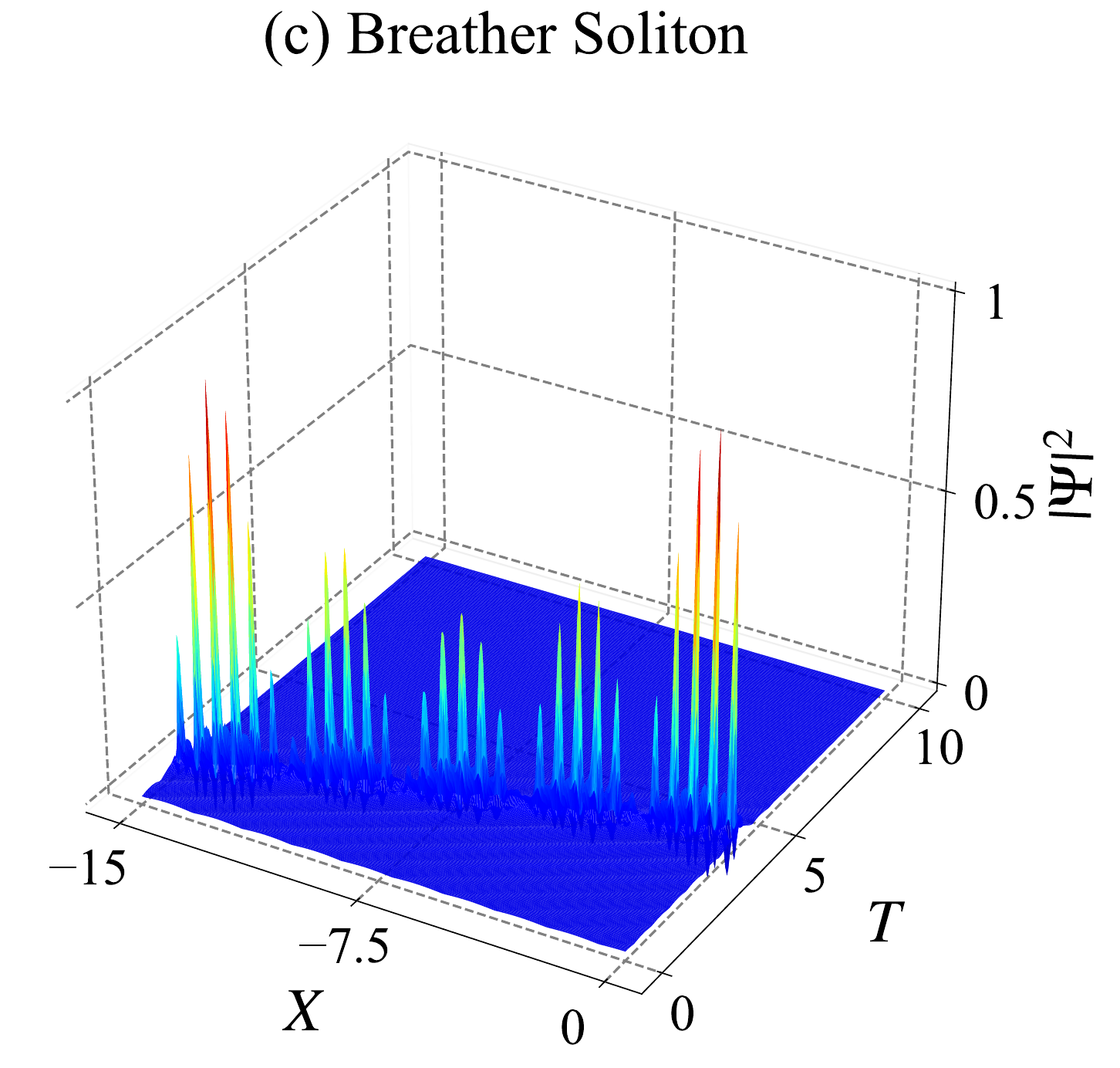}
}
\caption{Spatiotemporal evolution of magnetoelastic soliton solutions of the
NLSE, shown as color maps of $|\Psi(X,T)|^2$ representing (a) bright soliton,
(b) dark soliton and 
(c) Kuznetsov-Ma breather soliton.
}
\label{fig6}
\end{figure*}
\section{Soliton solutions}
\subsection{Bright soliton solution}
In this section, we present exact soliton solutions of the NLSE derived in Sec.~IV. Bright solitons arise in the focusing regime of the NLSE for $\mathcal{PQ} > 0$, where nonlinear self-focusing compensates dispersive spreading. We consider a traveling-wave solution of the form
\begin{equation}
\Psi(X,T)=\Psi_0\,\text{sech}[\kappa(X-vT)]
\exp\left[i(\Phi X-\Omega T)\right],
\label{eq62}
\end{equation}
where $\Psi_0$ denotes the soliton amplitude, $\kappa$ its inverse width, $v$ the soliton velocity, and $\Phi$ and $\Omega$ are the envelope wave number and nonlinear frequency shift, respectively. Introducing the comoving coordinate $z=X-vT$, the envelope becomes stationary in the moving frame.

Substituting Eq.~(\ref{eq62}) into Eq.~(\ref{eq59}) and separating real and imaginary parts, the imaginary terms vanish provided
by the velocity condition $v=2\mathcal{P}\Phi$, which identifies the soliton velocity with the group velocity of the carrier wave packet. The real part yields the relations
$\Omega=\mathcal{P}\kappa^2-\mathcal{Q}\Psi_0^2$ ,
together with the self-consistency condition $\kappa\equiv\Psi_0\sqrt{\frac{\mathcal{Q}}{\mathcal{P}}}. $ Choosing the soliton rest frame ($\Phi=0$), the bright soliton solution can be written in the canonical form
\begin{equation}
\Psi(X,T)
=\Psi_0\,\sec\text{h}\left(\kappa\,X\right)
\exp\left(i\kappa^2T\right).
\label{eq63}
\end{equation}

Restoring the original physical variables and the transverse magnetization component can be written as
\begin{equation}
m_x(x,t)
=
2\epsilon\Psi_0\,
\sec\text{h}\left[\epsilon\kappa(x-v_g t)\right]
\cos\left(k_0x-\omega_0t+\kappa^2\epsilon^2 t\right),
\label{eq64}
\end{equation}
representing a nondispersive, spatially localized magnetoelastic spin-wave packet propagating at the group velocity $v_g$.

The bright soliton solution of Eqs.~(\ref{eq63}) and (\ref{eq64}) shows that the magnetoelastic medium supports stable, nondispersive spin-wave packets when the focusing condition $\mathcal{P}\mathcal{Q}>0$ is satisfied. In this regime, the nonlinear frequency shift induced by magnetoelastic coupling compensates the intrinsic group-velocity dispersion of the exchange dominated magnon band, resulting in a self-localized excitation propagating at the group velocity $v_g$ without distortion. The soliton width is determined by the ratio $\mathcal{P}/\mathcal{Q}$, reflecting the balance between exchange driven dispersion and elastic field mediated nonlinearity.

\subsection{Dark soliton solution}
In the defocusing regime $\mathcal{P}\mathcal{Q}<0$, the NLSE supports dark
solitons corresponding to localized intensity dips on a finite-amplitude
background. Although the present exchange-dominated magnetoelastic model yields 
$\mathcal{P}\mathcal{Q}>0$ for the parameters considered above, 
we present the defocusing solution for completeness, since additional 
interactions like anisotropy or dipolar terms may modify the effective 
nonlinearity. Defining $\kappa'\equiv\Psi_0\sqrt{\left|\frac{\mathcal{Q}}{\mathcal{P}}\right|}$,
the dark soliton solution is
\begin{equation}
\Psi(X,T)=
\Psi_0\tanh(\kappa' X)\,
\exp\left(i\kappa'^2 T\right)
\label{eq65}
\end{equation}
Transforming back to physical variables gives
\begin{equation}
m_x(x,t)=
2\epsilon\Psi_0
\tanh\left[\epsilon\kappa'(x-v_g t)\right]
\cos\left(k_0x-\omega_0t+\kappa'^2\epsilon^2 t\right)
\label{eq66}
\end{equation}
describing a propagating magnetoelastic dark soliton traveling at the group
velocity $v_g$. In this regime, the nonlinear frequency shift counteracts the dispersive spreading in such a way that a localized depletion of the transverse magnetization remains stable during propagation. The soliton width here reflects the balance between exchange-driven dispersion and elastic-field–mediated nonlinearity.

\subsection{Breather soliton: Kuznetsov-Ma (KM) solution}
Beyond stationary solitons, the focusing NLSE $(\mathcal{P}\mathcal{Q}>0)$ also admits breather solutions associated with modulation instability. A representative
Kuznetsov-Ma (KM) breather is
\begin{equation}
\Psi(X,T)=
\Psi_0
\frac{\cosh(\Omega_b T+i\phi_0)+2\cos(KX)}
{\cosh(\Omega_b T)-2\cos(KX)}
\exp\left(i\kappa^2 T\right),
\label{eq67}
\end{equation}
where the breather frequency satisfies
\begin{equation}
\Omega_b^2=2\mathcal{Q}\Psi_0^2-\mathcal{P}K^2.
\label{eq68}
\end{equation}
In physical variables, the magnetization component becomes 
\begin{align} 
m_x(x,t) &= 2\epsilon\Psi_0\Re\Bigg\{ \frac{\cosh(\Omega_b\epsilon^2 t+i\phi_0) +2\cos[K\epsilon(x-v_g t)]} {\cosh(\Omega_b\epsilon^2 t) -2\cos[K\epsilon(x-v_g t)]} \nonumber\\ &\times \exp\!\left[i\left(k_0x-\omega_0t+\kappa^2\epsilon^2 t\right)\right] \Bigg\}, 
\label{eq69} 
\end{align}
this corresponds to a spatially localized spin-wave
packet whose envelope oscillates in time on the slow scale $\epsilon^2$.
Unlike stationary solitons, the breather represents a time-dependent bound
state arising from the interplay between dispersion and magnetoelastic
nonlinearity.

\begin{figure*}[t]
\centerline
\centerline{ 
\hspace{-0.5mm}
\includegraphics[scale=0.28]
{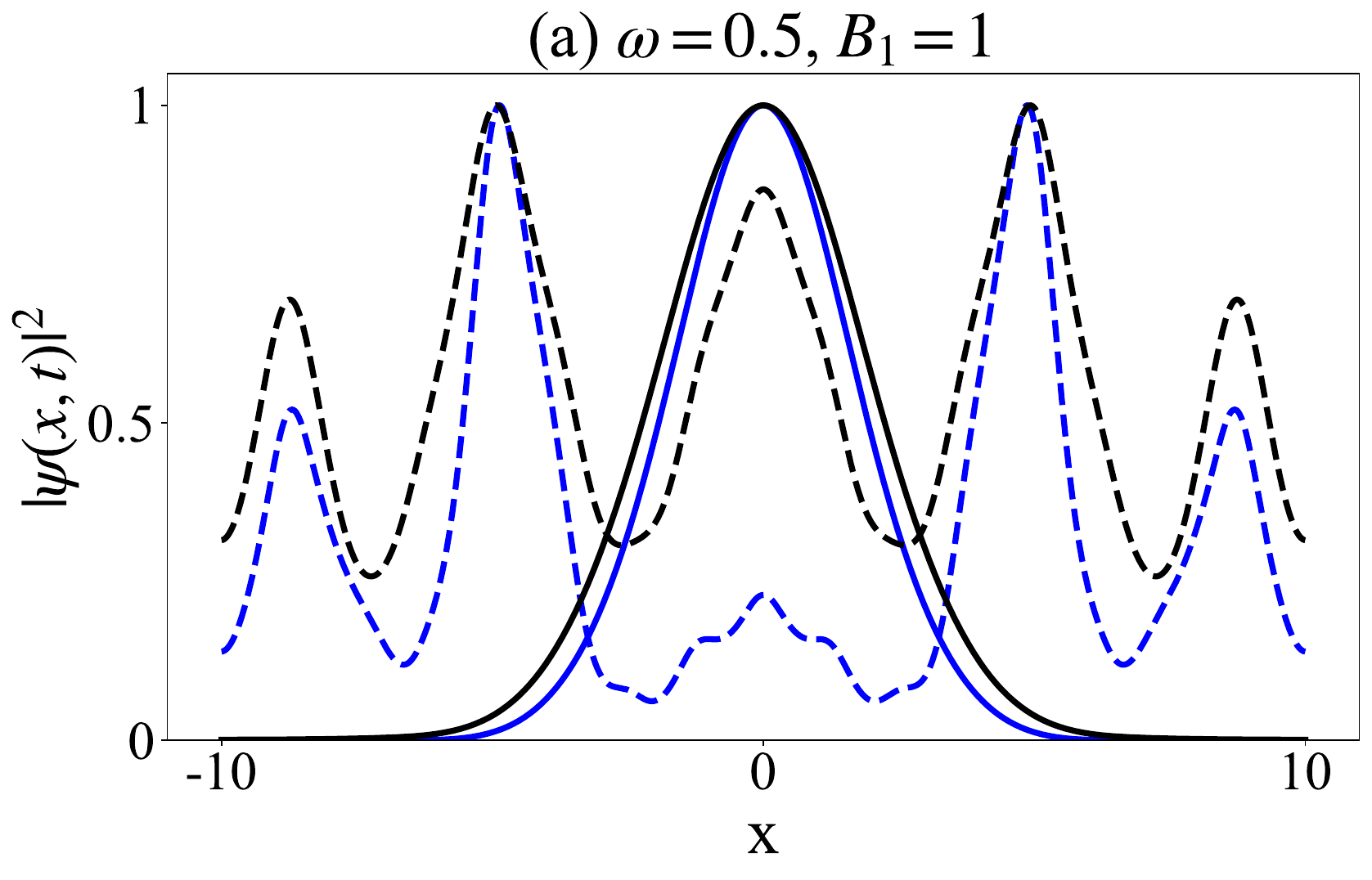}
\includegraphics[scale=0.28]{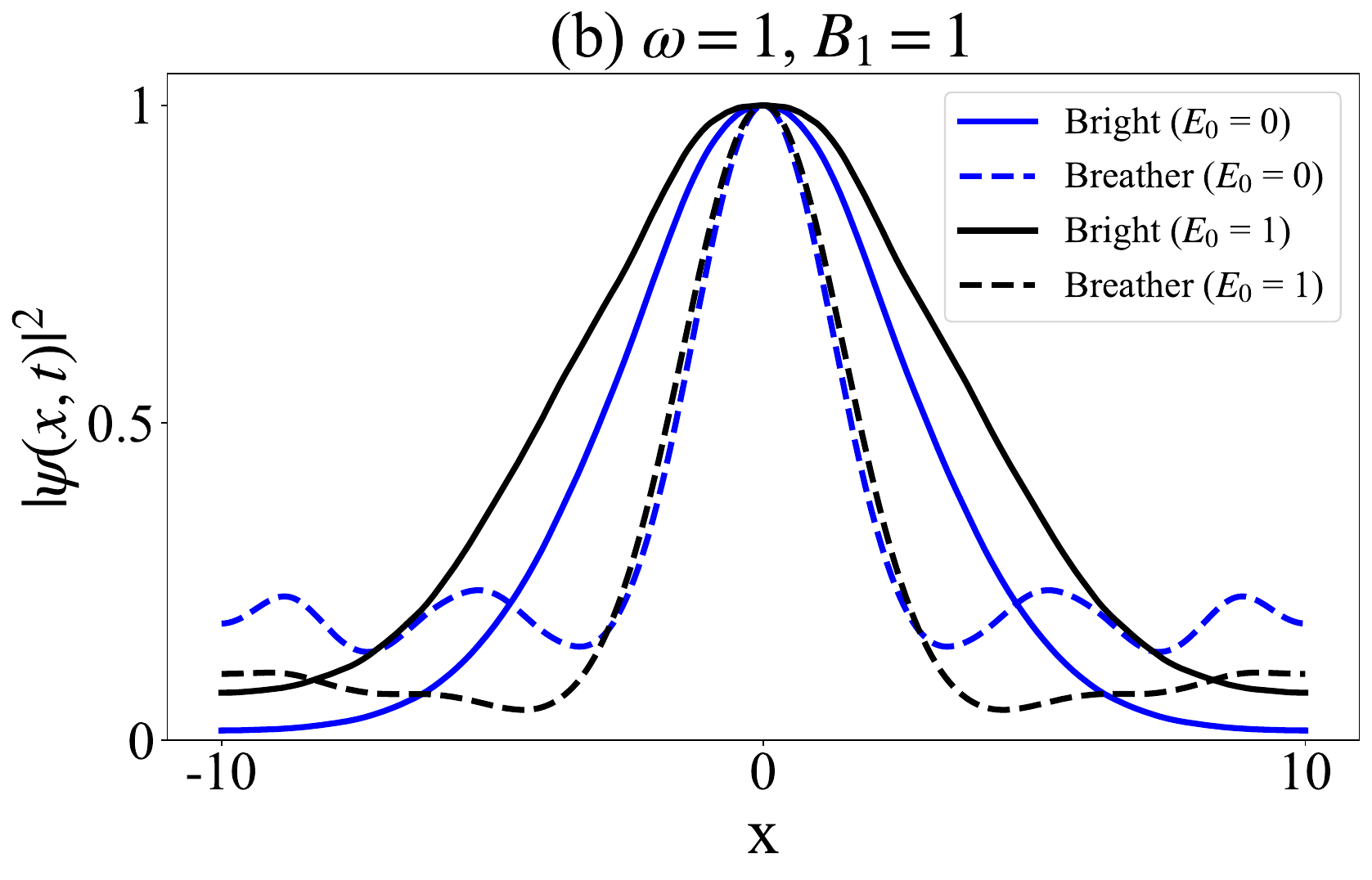}}
\centerline
\centerline{ 
\hspace{-0.5mm}
\includegraphics[scale=0.28]{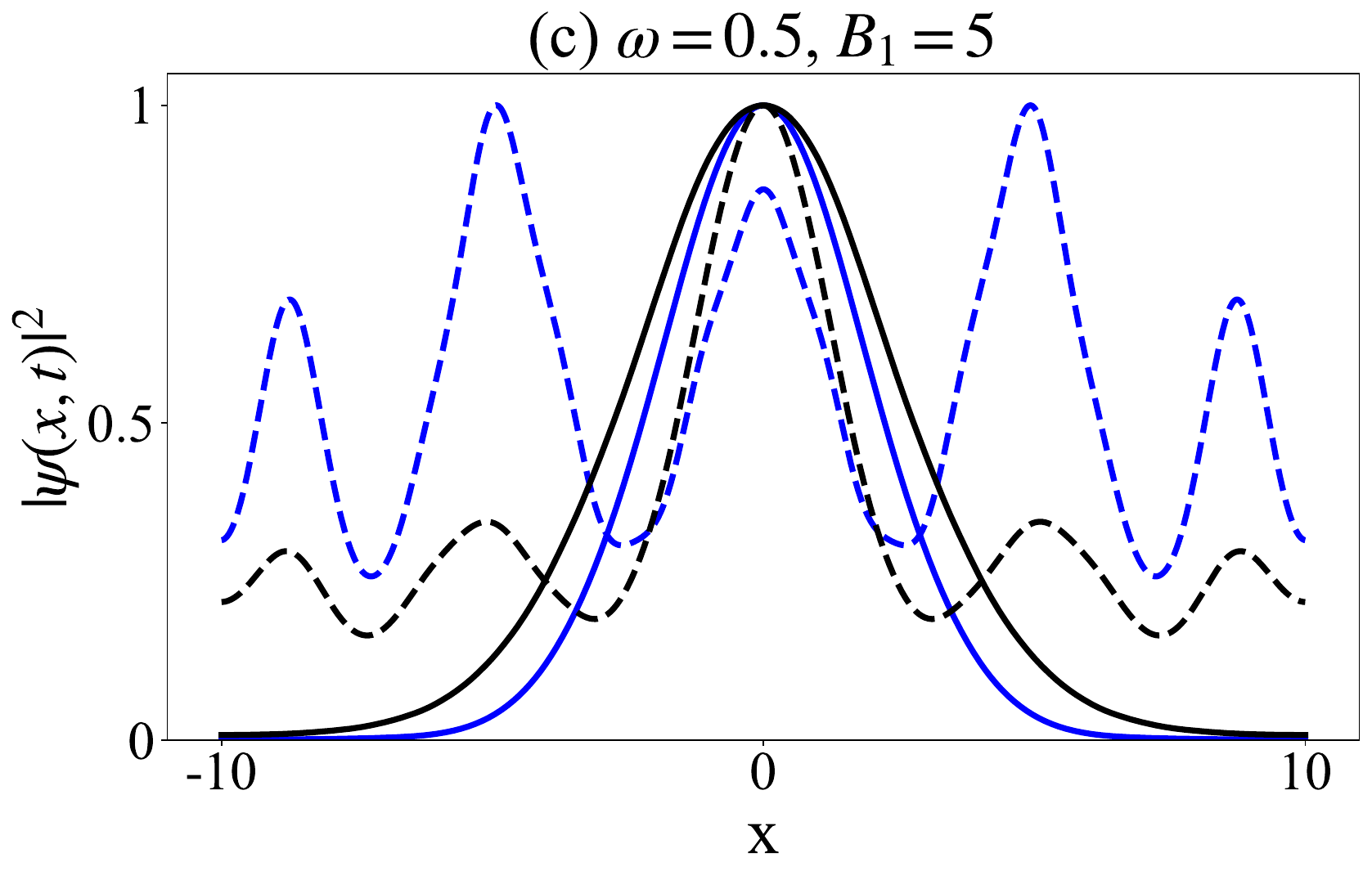}
\includegraphics[scale=0.28]{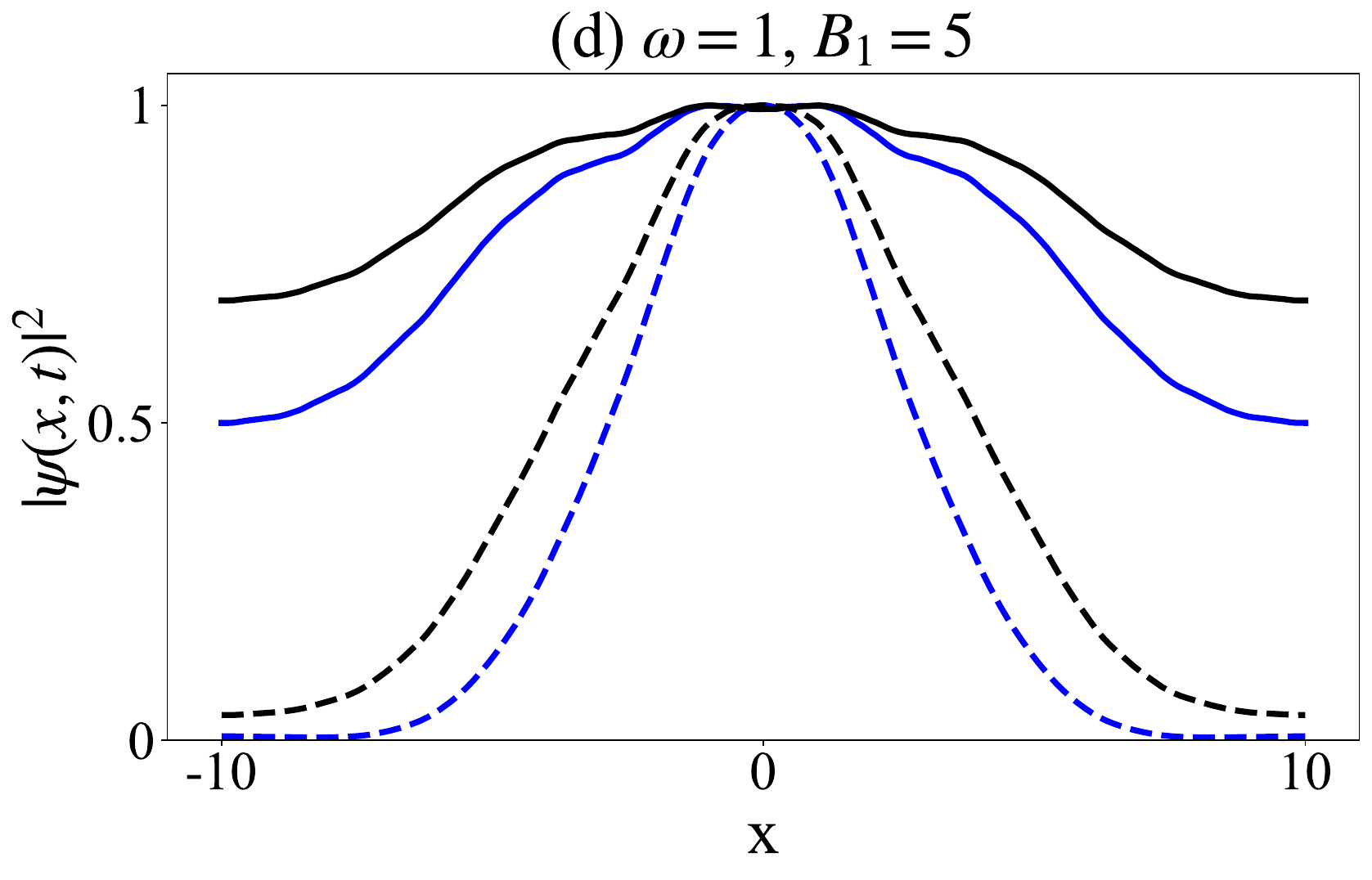}}
\caption{Squared envelope profiles $|\psi(x,0)|^{2}$ of nonlinear magnetoelastic excitations  for $\mathcal{P}\mathcal{Q}>0$ regime for different applied field frequencies $\omega$ and magnetoelastic coupling strengths $B_{1}$: (a) $\omega=0.5$, $B_{1}=1$, (b) $\omega=1$, $B_{1}=1$, (c) $\omega=0.5$, $B_{1}=5$, and (d) $\omega=1$, $B_{1}=5$. Solid lines denote bright soliton solutions while dashed lines corresponds to Kuznetsov-Ma breather solutions. Blue (black) curves correspond to $E_{0}=0$ ($E_{0}=1$). }
\label{fig7}
\end{figure*}
Fig.~\ref{fig6} shows representative spatiotemporal profiles of the
magnetoelastic soliton solutions of the NLSE derived in Sec.~IV C.
The color maps display the squared envelope amplitude $|\Psi(X,T)|^2$,
proportional to the intensity of the transverse magnetization modulation
within the reduced description.
Fig.~\ref{fig6}(a) presents a bright soliton in the focusing regime
$\mathcal{P}\mathcal{Q}>0$. The excitation remains spatially localized
and propagates without distortion, reflecting the balance between
exchange-driven dispersion and magnetoelastic nonlinearity.
The dark soliton solution in the defocusing regime
$\mathcal{P}\mathcal{Q}<0$ is shown in Fig.~\ref{fig6}(b).
Here the excitation appears as a localized intensity dip on a finite
background, preserving its shape during propagation due to nonlinear
phase modulation.
Fig.~\ref{fig6}(c) displays a KM breather supported by the
focusing NLSE. In contrast to stationary solitons, the breather exhibits
temporal oscillations of its amplitude while remaining spatially
localized, consistent with modulation instability dynamics. Overall, Fig.~\ref{fig6} demonstrates that the magnetoelastic NLSE supports both stationary and intrinsically time-dependent localized nonlinear excitations arising from the interplay of dispersion and nonlinearity.

Fig.~\ref{fig7} presents the spatial profiles $|\psi(x,0)|^2$ of nonlinear
magnetoelastic excitations obtained from the focusing NLSE
($\mathcal{P}\mathcal{Q}>0$), for different applied field frequencies
$\omega$ and magnetoelastic coupling strengths $B_1$. In all cases, the bright soliton exhibits a symmetric localized
$\mathrm{sech}^2$-type intensity profile, reflecting the exact
compensation between magnon dispersion,
originating from exchange stiffness and dipolar contributions,
and the cubic nonlinearity $\mathcal{Q}$ induced by magnetoelastic
coupling. Increasing $B_1$ enhances the nonlinear frequency shift
of the spin-wave packet, leading to a systematic increase in peak
amplitude accompanied by spatial compression of the envelope
as seen from  Figs.~\ref{fig7}(a)–\ref{fig7}(d).
This behavior follows from the NLSE amplitude width constraint
$A_{\mathrm{max}}^{2} L^{2} \propto \mathcal{P}/\mathcal{Q}$. Stronger nonlinearity increases self-focusing of the transverse
magnetization and reduces the soliton width in order to preserve
the dispersion nonlinearity balance. The KM breather solutions arise from modulation instability of a finite amplitude carrier. For $E_0=1$ the breather exhibits central amplification relative to the background while
remaining spatially localized. The weak oscillatory characteristics
originate from interference between the carrier and unstable
sideband modes selected by the magnetoelastic dispersion relation.
With increasing $B_1$, the KM breather becomes more localized and its
amplification increases, indicating enhancement of the modulation
instability growth rate.

The influence of the applied field frequency $\omega$ is more subtle and primarily enters through the effective dispersion coefficient $\mathcal{P}$ through the curvature of the
magnon band. Comparing Figs.~\ref{fig7}(a) and ~\ref{fig7}(b) at fixed $B_1=1$, and panels Figs.~\ref{fig7}(c) and ~\ref{fig7}(d) at fixed $B_1=5$, we found that a changing $\omega$ produces moderate variations in envelope width without altering the qualitative structure of the solutions. This confirms that $\omega$ tunes the dispersive length scale of the spin-wave packet, whereas $B_1$ controls the nonlinear localization strength. 
The absence of radiative tails confirms that these excitations are
robust nonlinear eigenmodes of the system.

\begin{figure}[t]
\vspace{-0.3cm}
\centerline{ 
\includegraphics[scale=0.34]{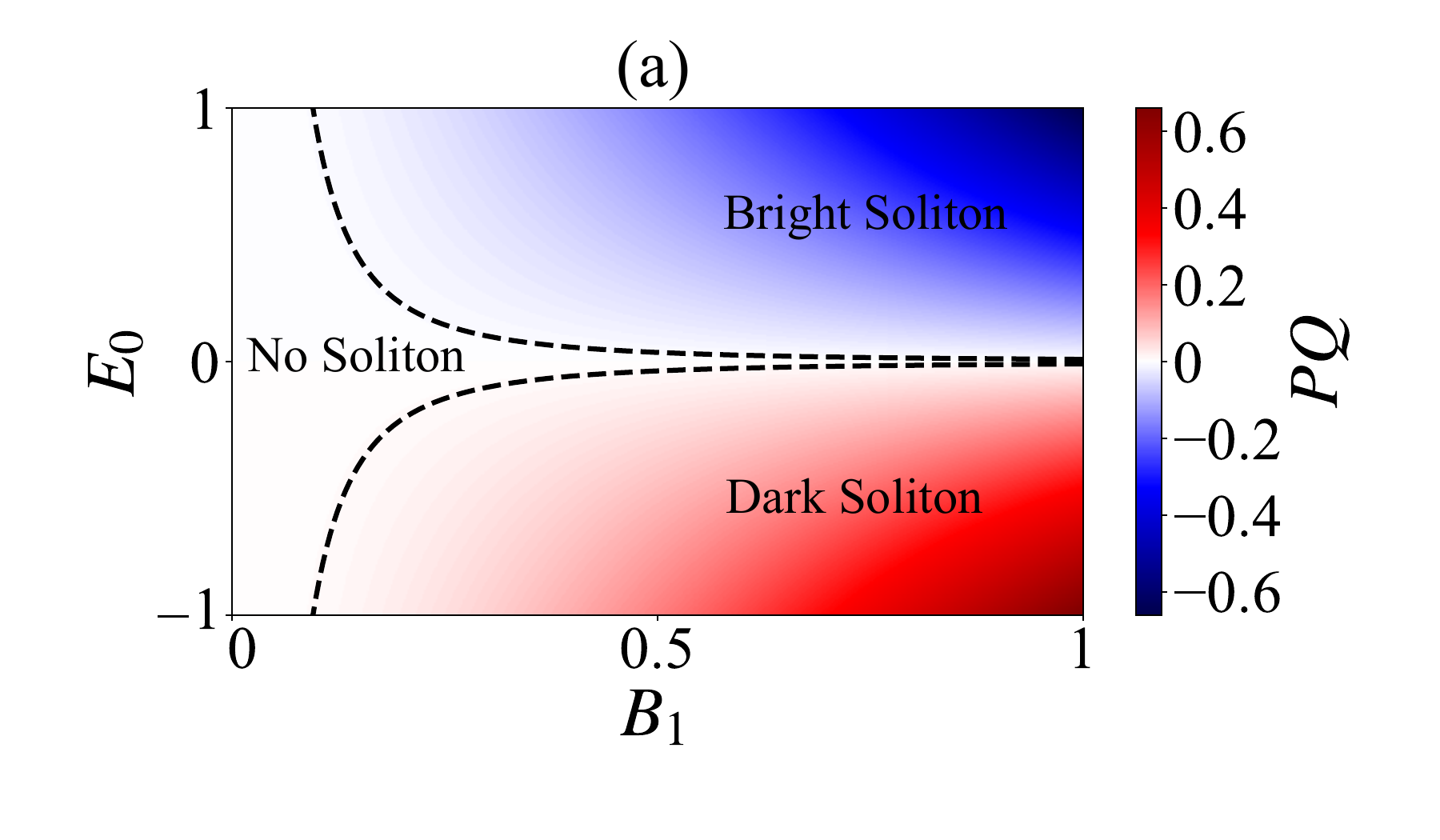}}
\vspace{-0.7cm}
\centerline{ 
\includegraphics[scale=0.34]{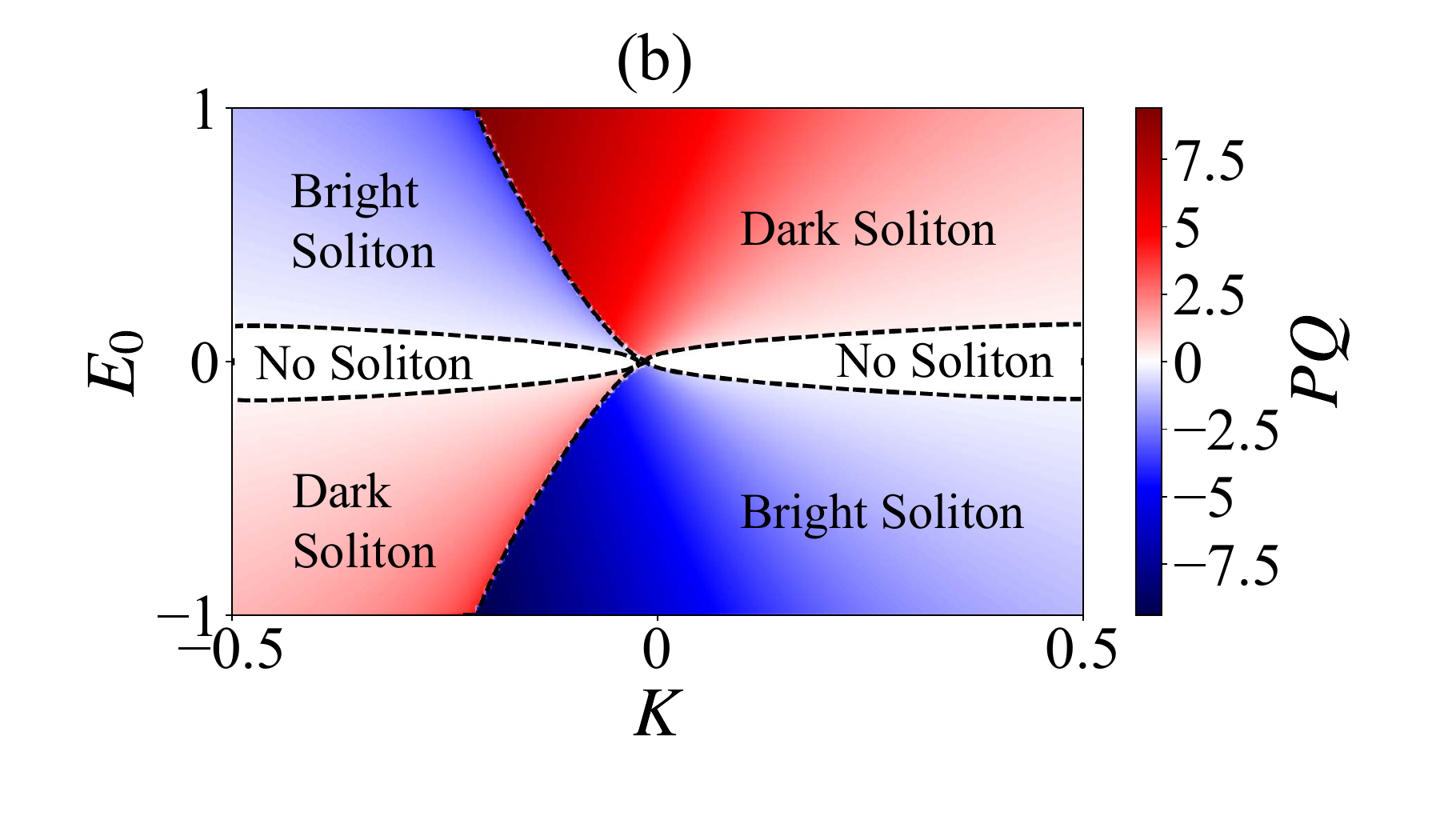}
}
\caption{
Phase diagrams showing the sign and magnitude of the product $\mathcal{P}\mathcal{Q}$ and the resulting soliton regimes. 
(Top) Phase diagram in the $(B_{1},E_{0})$ plane and 
(bottom) phase diagram in the $(K,E_{0})$ plane. 
Dashed contours indicate $\mathcal{P}\mathcal{Q}=0$. 
Regions labeled bright and dark soliton correspond to the focusing and defocusing regimes, respectively.
}
\label{fig8}
\end{figure}

The phase diagram shown in Fig.~\ref{fig8} maps the effective nonlinear-dispersive regime of the reduced magnetoelastic NLSE through the product $\mathcal{P}\mathcal{Q}$. It illustrates the role of $E_{0}$, $B_{1}$, and $K$ jointly to determine the nature of soliton solutions. It is observed that with the increase in $B_{1}$ monotonically increase the effective magnetoelastic nonlinearity as seen from Fig.~\ref{fig8}(a). and It drives the system away from the neutral line into robust soliton regimes, with the color scale indicating the magnitude of $\mathcal{P}\mathcal{Q}$; a finite background $E_{0}$ shifts the operating point and opens an extended parameter window for localized states by modifying the nonlinear frequency shift. Fig.~\ref{fig8}(b) gives the phase diagram of $\mathcal{PQ}$ with $E_0$ and $K$. It is observed that the phase diagram has a fourfold wedge structure in which bright and dark soliton domains interchange across $K=0$ and across $E_{0}$-dependent neutral lines. The large values of $K$ and $E_{0}$ can enhance the value of $\mathcal{P}\mathcal{Q}$ and thus stabilize the corresponding soliton solutions. Physically, these phase diagrams indicate that (i) the magnetoelastic coupling provides a direct control parameter for tuning between no-soliton, bright-soliton, and dark-soliton phases and (ii) a finite carrier or background field qualitatively modifies the nonlinearity and enables breather-type excitations on a finite background.

\begin{figure*}[hbt]
\centerline{ 
\includegraphics[scale=0.25
]{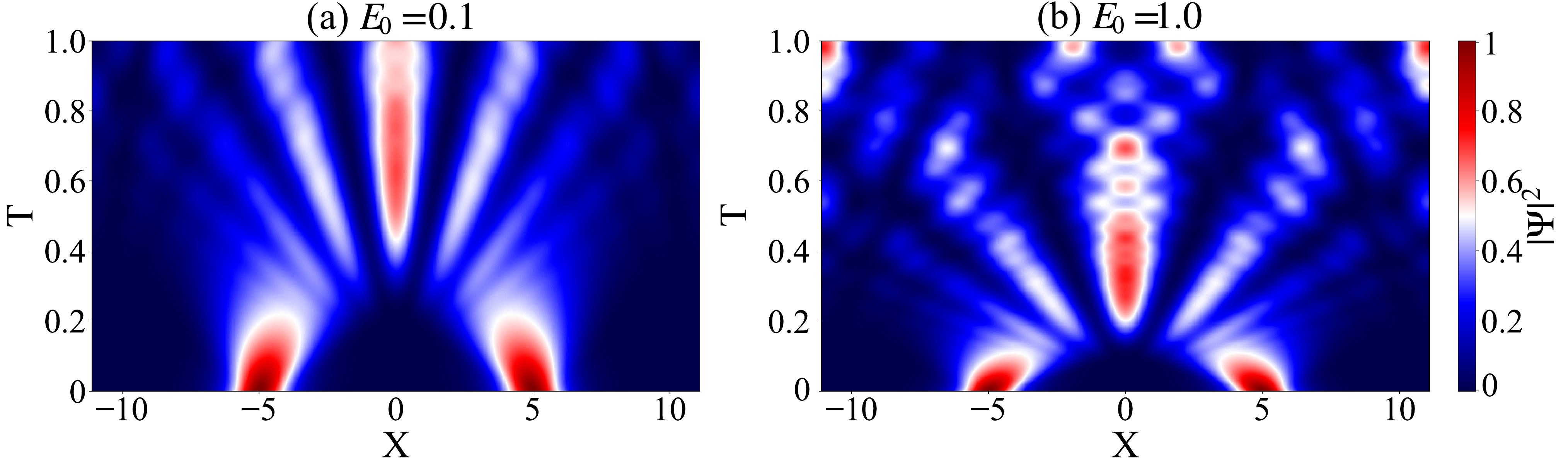}
}
\caption{Space–time maps of the envelope intensity $|\psi(x,t)|^{2}$ showing soliton interference patterns for (a) $E_{0}=0.1$ and (b) $E_{0}=1.0$.}
\label{fig9}
\end{figure*}

\subsection{Soliton interference dynamics}
To investigate the soliton interference effects, the initial condition is chosen as a coherent superposition of two counter propagating localized wave packets on a uniform background field $E_0$
\begin{equation}
\psi(X,0)=
\psi_1(X)+\psi_2(X)+E_0 ,
\label{eq70}
\end{equation}
where, $\psi_j(X)$ are defined as
\begin{equation}
\psi_j(X)=
\psi_0\,\mathrm{sech}\left[\kappa(X-X_j)\right]
\exp\left(i k_j X + i\phi_j\right),
\label{eq71}
\end{equation}
where $j=1,2$ and $\psi_0$ is the soliton amplitude,  $k_j$ the carrier wave number, and $\phi_j$ the initial phase. The soliton envelope intensity is given by
\begin{equation}
I(X,T)=|\psi(X,T)|^2 ,
\label{eq72}
\end{equation}
The space-time maps of the envelop intensity is presented in Fig.~\ref{fig9}, considering both linear interference and nonlinear interaction effects. The cubic nonlinearity in Eq.~(\ref{eq59}), contributes a local intensity dependent phase accumulation approximately given by
\begin{equation}
\delta(X,T)
=
\mathcal{Q}\int_0^T |\psi(X,T')|^2\,dT' ,
\label{eq73}
\end{equation}
leading to self and cross-phase modulation between the interfering solitons.

For $E_0=0.1$, the interaction remains nearly elastic and the interference pattern is dominated by coherent superposition, producing the characteristic X-shaped fringe structure of soliton collisions as seen from Fig.~\ref{fig9}(a).  For $E_0=1.0$, the dynamics are qualitatively modified by background induced modulation instability as seen from Fig.~\ref{fig9}(b).
Linear stability analysis of a plane-wave state $\psi = E_0 e^{-i\mathcal{Q}E_0^2 T}$ results the dispersion relation
\begin{equation}
\Omega^2(K)
=
\mathcal{P}^2 K^4
-2\mathcal{P}\mathcal{Q}E_0^2 K^2 
\label{eq74}
\end{equation}
so that modes satisfying
$K^2 < 2\mathcal{Q}E_0^2/\mathcal{P}$
become unstable in the focusing regime
$\mathcal{P}\mathcal{Q}>0$.
During soliton overlap, this instability enhances nonlinear phase
accumulation and temporarily transfers energy from the background
to localized envelope modes. As a result, the interference
pattern exhibits stronger central localization, reduced fringe spacing,
and breather-like oscillatory structures. 
Since both the nonlinear coefficient 
$\mathcal{Q}$ and the effective dispersion $\mathcal{P}$ depend parametrically on the applied electric field strength $E_0$, the modulation instability threshold and collision dynamics are electrically tunable. 
In the weak-background limit the dynamics
approach elastic soliton scattering, where as finite background
fields activate modulation instability and promote transient
breather assisted energy redistribution. The interference patterns
therefore provide a direct probe of nonlinear coherence and
stability in propagating magnetoelastic spin-wave packets.

\section{Conclusion}
In this work, we study the nonlinear magnetoelastic wave dynamics and electrically tunable soliton excitations in hexagonal multiferroic media. By systematically varying the magnetoelastic coupling strength, we demonstrate a transition from weakly nonlinear, nearly periodic oscillations to strongly anharmonic yet phase-coherent multimode dynamics. Our results suggest that despite the increasing nonlinearity, the dynamics remain bounded and regular evolving towards distorted limit-cycle behavior rather than chaotic motion. The excitation spectra and dispersion relations reveal that this nonlinear evolution originates from strong magnon-phonon hybridization and coupling induced renormalization of the collective excitation branches. As a result, the magnetic, elastic, and polarization subsystems exhibit synchronized nonlinear oscillations and coherent energy redistribution mediated by magnetoelastic interactions. Moreover, the coupled dynamics can be reduced to an effective nonlinear Schr\"{o}dinger equation that supports localized nonlinear excitations, including bright and dark and Kuznetsov-Ma breather like solitons. It is observed that the applied electric field modifies both the effective nonlinear coefficient and the dispersion curvature, thereby enabling continuous tuning of soliton amplitude, width, energy, and stability. In particular we found that the electric field induce saddle-node bifurcation in the magnetization phase space.  This result establishes a critical threshold separating multistable and monostable regimes, which directly governs the existence and robustness of localized solutions. The resulting phase diagrams map the nonlinear dispersive landscape of the system and identify electrically tunable domains of bright and dark solitons. Furthermore, the interference dynamics of counter propagating solitons reveal that a finite background field can trigger modulation instability. This leads to breather assisted energy localization and enables electric field control of the soliton collision dynamics. Overall, these results demonstrate that magnetoelastic and magnetoelectric couplings provide a powerful mechanism for coherent nonlinear control of spin-lattice excitations in multiferroic media. This work establishes a theoretical basis for electrically tunable soliton engineering and nonlinear magnon-phonon control in hexagonal multiferroic media.

\end{document}